\documentclass[%
 reprint,
 amsmath,amssymb,
 aps,
]{revtex4-1}

\usepackage{graphicx}% Include figure files
\usepackage{dcolumn}% Align table columns on decimal point
\usepackage{bm}% bold math
\newcommand{\beq}{\begin{equation}}
\newcommand{\eeq}{\end{equation}}
\newcommand{\beqa}{\begin{eqnarray}}
\newcommand{\eeqa}{\end{eqnarray}}

\newcommand{\kBT}{\mbox{$k_{\rm B}T$}}

\renewcommand{\vec}[1]{\mbox{\boldmath$#1$}}

\newcommand{\simless}{\stackrel{<}{\sim}}

\begin{document}

\preprint{APS/123-QED}

\title{
Height of a faceted macrostep for sticky steps in a step-faceting zone 
%Height of faceted macrosteps for sticky steps in
%non-equilibrium steady-state in step-faceting zone 
%Driving force dependence of the profile of a faceted macrostep for sticky steps:
%non-equilibrium steady-state in step-faceting zone 
%Driving force dependence of the profile of a faceted macrostep in step-faceting zone in non-equilibrium steady-state
%Attachment-detachment of steps with a faceted macrostep for sticky steps: non-equilibrium steady-state in step-faceting zone
%Non-equilibrium steady state of faceted macrosteps in the step-faceting zone
%Dynamics in Non-Equilibrium Steady-State on a Vicinal Surface with a Faceted Macrostep in Step-Faceting Zone
%Driving force dependence of step detachments from a faceted macrostep in step-faceting zone in non-equilibrium steady-state
%Height of a Faceted Macrostep in Step-Faceting Zone around Non-Equilibrium Steady-State
%Driving force dependence of 
}

\author{Noriko Akutsu}
 \email{nori3@phys.osakac.ac.jp, nori@phys.osakac.ac.jp}
\affiliation{%
 Faculty of Engineering, Osaka Electro-Communication University, Hatsu-cho, Neyagawa, Osaka 572-8530, Japan
}%


%\affiliation{
%\affiliation{
%\affiliation{%


\date{\today}% It is always \today, today,

\begin{abstract}
The driving force dependence of the surface velocity and the average height of faceted merged steps, the terrace-surface-slope, and the elementary step velocity in the non-equilibrium steady-state are studied using the Monte Carlo method.
The Monte Carlo study is based on a lattice model, the restricted solid-on-solid model with point-contact type step--step attraction (p-RSOS model).
The temperature is selected to be in the step-faceting zone where the surface is surrounded by the (001) terrace and the (111) faceted step at equilibrium.
Long time simulations are performed at this temperature to obtain steady-states for the different driving forces that influence the growth/recession of the surface.
A Wulff figure of the p-RSOS model is produced through the anomalous surface tension calculated using the density-matrix renormalization group method.
The characteristics of the faceted macrostep profile at equilibrium are classified with respect to the connectivity of the surface tension.
This surface tension connectivity also leads to a faceting diagram, where the separated areas are respectively classified as a Gruber-Mullins-Pokrovsky-Talapov zone, step droplet zone, and step-faceting zone.
Although the p-RSOS model is a simplified model, the model shows a wide variety of dynamics in the step-faceting zone.
There are four characteristic driving forces, $\Delta \mu_y$, $\Delta \mu_f$, $\Delta \mu_{co}$, and $\Delta \mu_R$.
For the absolute value of the driving force, $|\Delta \mu|$ is smaller than Max[$\Delta \mu_y, \Delta \mu_f]$, the step attachment-detachments are inhibited, and the vicinal surface consists of (001) terraces and the (111) side surfaces of the faceted macrosteps.
For Max[$\Delta \mu_y, \Delta \mu_f]<|\Delta \mu|<\Delta \mu_{co}$, the surface grows/recedes intermittently through the two-dimensional (2D) heterogeneous nucleation at the facet edge of the macrostep.
For $\Delta \mu_{co}<|\Delta \mu|<\Delta \mu_R$, the surface grows/recedes with the successive attachment-detachment of steps to/from a macrostep.
When $|\Delta \mu|$ exceeds $\Delta \mu_R $, the macrostep vanishes and the surface roughens kinetically.
Classical 2D heterogeneous multi-nucleation was determined to be valid with slight modifications based on the Monte Carlo results of the step velocity and the change in the surface slope of the ``terrace''.
The finite size effects were also determined to be distinctive near equilibrium. 
\begin{description}
\item[PACS numbers]
% \verb+ 
81.10.Aj 64.60.Q- 82.60.Nh 68.35.Md 02.70.Uu 81.10.Dn 68.35.Ct 05.70.Np %+
\end{description}
\end{abstract}

%\pacs{81.10.Aj 64.60.Q- 82.60.Nh 68.35.Md 02.70.Uu 81.10.Dn 68.35.Ct 05.70.Np}
%nucleation 64.60.Q- 68.55.A-(thin film) 82.60.Nh(chemistry) 
%Monte Carlo 02.70.Uu 64.60.De 
%Interface thermodynamics 05.70.Np 68.35.Ct(roughness) 68.35.Md(statphys)
%Crystal growth 81.10.Aj 81.10.Dn(solution growth)
%solid surfaces, solid-solid interfaces 68.35.-p

\maketitle

%\tableofcontents

\section{Introduction}

%(4H-SiC)
Faceted macrosteps have sometimes been considered to degrade the quality of grown crystals.
In the case of solution growth for 4H-SiC \cite{mitani} as an example, which is expected to be used for future power devices, the faceted macrosteps near equilibrium hinder the preparation of good quality crystals that satisfy the requirements for electrical devices.
Therefore, to control the dynamics of macrosteps, the fundamentals regarding the formation of macrosteps should be clarified.

%planar surface is singular
%singular planar surface 
For smooth surfaces, the approaches based on the time-dependent Ginzburg-Landau (TDGL) equation for surface motion \cite{burkhardt, enomoto, barabasi, saito78, pimpinelli}, which is used to study rough surfaces, are not valid.
The nucleation model is instead known to be more effective.
In Saito's solution for the TDGL equation of a smooth surface with a modified discrete Gaussian (MDG) model \cite{saito78}, the two-dimensional (2D) nucleus is not included.
Based on the TDGL equation of the surface, the surface cannot grow until $\Delta \mu$ exceeds a ``spinodal'' value, $\Delta \mu_c$.
Here, $\Delta \mu=\mu_{\rm ambient}-\mu_{\rm crystal}$ is the driving force, where $\mu_{\rm crystal}$ is the chemical potential of the bulk crystal and $\mu_{\rm ambient}$ is the chemical potential of the ambient phase.
This suggests that the excitation of islands on the surface, which can be 2D nuclei, corresponds to a higher order response to the driving force with respect to the crystal growth.

The phase field method \cite{phase_field} is also known as a powerful tool to study the solidifications or other non-equilibrium phenomena accompanied by phase changes.
In the phase field method, the phase boundary is assumed to be analytic and differentiable.
The faceted structure on the surface can be simulated with the phase field method by the introduction of a strong anisotropy to the interface tension.
The planes can be imitated by curved surfaces with small curvature.
 However, this slight difference in the curvature causes a significant difference in the long time behavior, {\it i.e.,} the non-equilibrium steady-state behavior.
For example, the singularity of the flat smooth interface inhibits the growth/recession of the crystal without 2D nucleation processes or screw dislocations \cite{bcf}.

With respect to nucleation, the nucleation model interprets how the large clusters (domains) are formed; therefore, the nucleation model is widely accepted for study of the dynamics around the first-order phase transitions \cite{bcf, pimpinelli, kashchiev, dubrovskii}.
However, for quantitative study, the classical nucleation theory \cite{becker, bcf,ookawa} often disagrees with the experimental observations or the results of large-scale molecular dynamics by an order of $10^3$ \cite{tanaka14}. 
Therefore, many improvements in the nucleation theory have been reported \cite{kashchiev, dubrovskii}. 
In recent years, the notion of modern ``multi-nucleation'' \cite{kimura12,kimura16,tanaka14,nanev15}, where the growing clusters change the crystal structure at a certain size during growth, has attracted attention and opened a new research area.

A vicinal surface with faceted macrosteps is surrounded by smooth surfaces; therefore, the surface is considered to grow/recede by way of 2D nucleation \cite{bcf, becker, ookawa}.
However, macrosteps are considered to be unstable at equilibrium \cite{k_roughening,uwaha92} without impurities or adatoms. 
Therefore, the dynamics of a vicinal surface with faceted macrosteps in the non-equilibrium steady-state have not been studied sufficiently.

%%%%% FIGURE  %%%%%

\begin{figure}%[h]%[htbp]
%\begin{center}
\centering
\includegraphics[width=8.0 cm,clip]{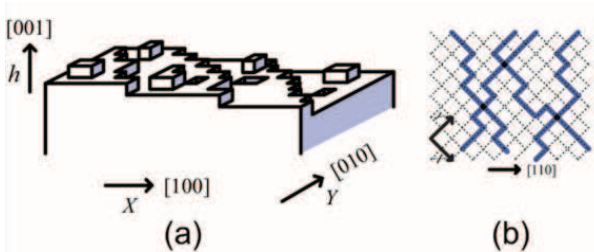}%
%\end{center}
\caption{
(a) Perspective view of the RSOS model tilted toward the $\langle 110\rangle$ direction. 
(b) Top-down view of the RSOS model.
Thick blue lines represent surface steps.
Reproduced from \cite{akutsu16-3}, with the permission of Hindawi Publishing Corporation. 
}  
\label{vicinal}
\end{figure}
%%%%% FIGURE  %%%%%

With respect to the stable faceted macrostep, we reported a study on a restricted solid-on-solid (RSOS) model with a point-contact-type step--step attraction (p-RSOS model, Fig.~\ref{vicinal}) \cite{akutsu09,akutsu11,akutsu11JPCM,akutsu12,akutsu14, akutsu16,akutsu16-2,akutsu16-3}. 
Here, ``restricted'' means that the height difference between nearest neighbor sites is restricted to $\{0,\pm1\}$. 
The origin of the point-contact-type step--step attraction is considered to be the orbital overlap of the dangling bonds at the meeting point of neighboring steps. 
The energy gained by the formation of a bonding state is regarded as the attractive energy between steps.

The characteristic of the p-RSOS model is the discontinuous surface tension at low temperatures \cite{akutsu11JPCM, akutsu12, akutsu14}.
Macrosteps are stabilized when the surface free energy has anomalous anisotropy \cite{cabrera, cabrera64}.
Therefore, it seems a simple task to calculate the surface free energy explicitly with a standard method of the statistical mechanics.
However, due to the large contribution of thermal fluctuations in a low-dimensional substance \cite{mermin}, it is difficult to obtain reliable results theoretically using the mean field approximation. 
Therefore, to obtain reliable results, the density-matrix renormalization group (DMRG) method \cite{dmrg,dmrg2,dmrg3,pwfrg,pwfrg2,pwfrg3} is used for calculation of the surface tension (the surface free energy per normal unit area).

\begin{figure}%[h]%[htbp]
%\begin{center}
\centering
\includegraphics[width=8.0 cm,clip]{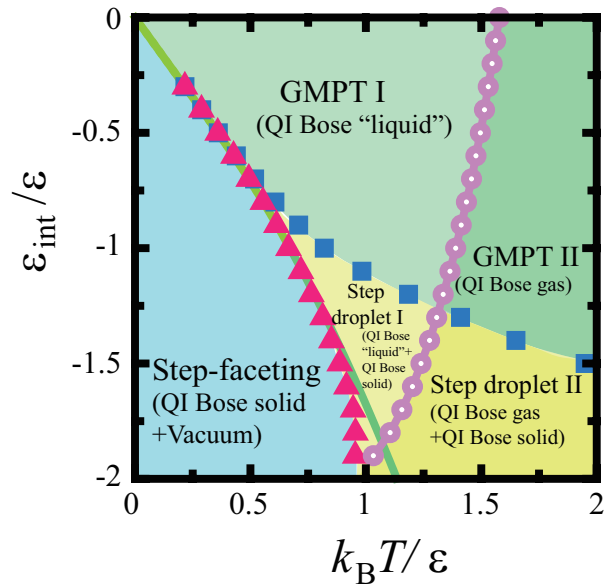}%
%\end{center}
\caption{
(Color online)
Faceting diagram of the p-RSOS model for a vicinal surface obtained using the DMRG method.
Squares: calculated values of $T_{f,1}$.  
Triangles: calculated values of $T_{f,2}$.  
Open circles: calculated roughening transition temperatures of the (001) surface.
Solid line: zone boundary line calculated using the 2D Ising model.
For definitions and details regarding the QI Bose solid, liquid, and gas, please refer to Akutsu \cite{akutsu16}. 
Reproduced from \cite{akutsu16}, with the permission of AIP Publishing. 
}  
\label{phase_diagram}
\end{figure}

A faceting diagram (Fig.~\ref{phase_diagram}) that corresponds to the connectivity of the surface tension was obtained \cite{akutsu16}.
The faceted macrostep is stabilized in the step faceting zone and in the step-droplet zone.
In the Gruber-Mullins-Pokrovsky-Talapov (GMPT) zone, there is no faceted macrostep; the vicinal surface obeys the GMPT universal behavior \cite{gmpt, gmpt2, beijeren87}.
In the step faceting zone, the vicinal surface consists of (001) terraces and a single (111) surface which forms the side surface of a faceted macrostep.
In contrast, the vicinal surface in the step droplet zone consists of a single (111) surface and surfaces with slope $p_1$.
The characteristic of the height profile of the faceted macrostep is also classified by the connectivity of the surface tension at equilibrium \cite{akutsu16-3}.

%%%%% FIGURE  %%%%%

%%%%% FIGURE  %%%%%

In this article, we study the driving force dependence of the surface velocity $V$, the average height of faceted merged steps $\langle n \rangle$, the terrace-surface-slope $p_1$, and the elementary step velocity $v_{\rm step}$ in the non-equilibrium steady-state using the Monte Carlo method for a vicinal surface with a faceted macrostep in the p-RSOS model.
The temperature is selected to be in the step-faceting zone where the surface is surrounded by a (001) terrace and (111) faceted step at equilibrium.
The Wulff figure of the p-RSOS model is produced from the anomalous surface tension calculated with the DMRG method.

To clarify the effect of the discontinuous surface tension on the dynamics of the surface, the following effects are excluded from our model: the surface diffusion \cite{uwaha92,sato2000}, the volume diffusion \cite{nishinaga89,nishinaga93}, the elastic interaction \cite{calogero, sutherland, alerhand, williams93, yamamoto94, song94, song95, dijken97, sudoh01, lossig96, bhattachariee, shenoy98, shenoy2000}, the long range attractive interaction \cite{jayaprakash84-2,tonchev16}, the Ehrlich-Schwoeble effects \cite{ehrlich,schwoeble}, and the effects caused by thermal expansion \cite{osman}.

This article is organized as follows.
In \S \ref{model}, the model Hamiltonian and the surface tension calculated with the DMRG method are shown.
In \S \ref{MCresults}, we present the results obtained from long time Monte Carlo simulation for an average-sized faceted macrostep $\langle n \rangle$ and the growth rate of the surface, $V$. 
In \S \ref{analysis}, the Monte Carlo results are analyzed based on the classical 2D heterogeneous nucleation and the classical 2D heterogeneous multi-nucleation.
The driving force dependence of the slope of the ``terrace'' and the step velocity are also presented.
The crossover to the kinetic roughened surface is discussed in \S \ref{k-roughening}.
Further discussions are given in \S \ref{discussion}, and conclusions are presented in \S \ref{conclusions}.

%\section{Restricted solid-on-solid model with point-contact-type step--step attraction}

%\subsection{Restricted solid-on-solid model with point-contact-type step--step attraction}

%\section{Results}

\section{The model\label{model}}

\subsection{The p-RSOS model}
%In this section, we will explain the discontinuity of surface tension.
The microscopic model considered in this study is the p-RSOS model (Fig.~\ref{vicinal}) \cite{akutsu09, akutsu11, akutsu11JPCM, akutsu12, akutsu14, akutsu16, akutsu16-2}. 
In this model, ``an atom'' corresponds to a unit cube. 
The Hamiltonian of the (001) surface can be written as
\beqa
&&{\cal H}_{\rm p-RSOS} = {\cal N}\epsilon_{\rm surf}+ \sum_{n,m} \epsilon 
[ |h(n+1,m)-h(n,m)|  \nonumber \\
&&
+|h(n,m+1)-h(n,m)|]   \nonumber \\
&& +\sum_{n,m} \epsilon_{\rm int}[ \delta(|h(n+1,m+1)-h(n,m)|,2)  \nonumber \\
&& 
+\delta(|h(n+1,m-1)-h(n,m)|,2)],   \label{hamil}
\eeqa
where ${\cal N}$ is the total number of lattice points, $\epsilon_{\rm surf}$ is the surface energy per unit cell on the planar (001) surface, $\epsilon$ is the microscopic step energy, $\delta(a,b)$ is the Kronecker delta, and $\epsilon_{\rm int}$ is the microscopic step--step interaction energy.
The summation with respect to $(n,m)$ is taken over all sites on the square lattice.
The RSOS condition is required implicitly.
When $\epsilon_{\rm int}$ is negative, the step--step interaction becomes attractive (sticky steps).

\subsection{Discontinuous surface tension}

The surface tension is the surface free energy per unit normal area.
The surface tension $\gamma_{\rm surf}(\vec{p})$ was calculated from the surface free energy $f(\vec{p})$ per projected $x $--$y$ area for the vicinal surface as
\beq
\gamma_{\rm surf}(\vec{p})= \frac{f(\vec{p})}{\sqrt{1+p_x^2+p_y^2}}, 
\eeq
where $\vec{p}= (p_x, p_y) $ is the surface gradient of the vicinal surface \cite{akutsu87}. 
The surface free energy $f(\vec{p})$ is calculated from the Andreev free energy \cite{andreev}, which is the grand potential on the grand partition function with respect to the number of steps (Appendix \ref{surface-tension}). 
The DMRG method was applied for calculation of the grand partition function.
The transfer matrix version of the DMRG method, which is known as the product wave function renormalization group (PWFRG) method \cite{pwfrg,pwfrg2,pwfrg3}, was used in this study.
Details of the method for calculation of the surface tension and the surface free energy are given in Appendix \ref{surface-tension}.

%%%%% FIGURE  %%%%%

\begin{figure}%[h]%[htbp]
%\begin{center}
\centering
\includegraphics[width=7.5 cm,clip]{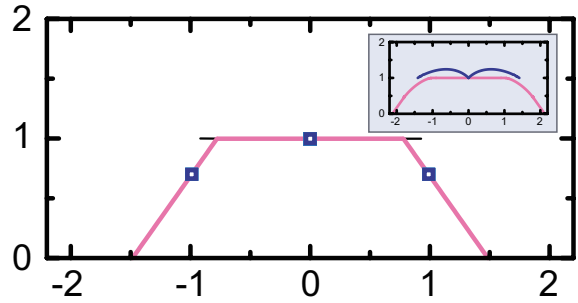}%
%\end{center}
\caption{
Polar graph of surface tension (Wulff figures \cite{laue}) and Andreev's free energies (ECSs) from 
DMRG calculations.
p-RSOS model in the step-faceting zone ($\epsilon_{\rm int}/\epsilon=-0.9$).
Filled squares: Surface tension $\gamma (\vec{p})/\epsilon$.
Thin black lines: The (001) surface in metastable states.
(Inset) Original RSOS model ($\epsilon_{\rm int}=0$).
$\kBT/\epsilon =0.4$.
Dark lines or filled squares: Surface tension $\gamma (\vec{p})/\epsilon$.
Pale lines: Andreev's free energy calculated using the DMRG method \cite{dmrg,dmrg2,dmrg3,pwfrg,pwfrg2,pwfrg3}.
$\epsilon_{\rm surf}$ is assumed to equal $\epsilon$.
}  
\label{gampolarT04}
\end{figure}
%%%%% FIGURE  %%%%%

The polar graphs of the surface tension are shown in Fig.~\ref{gampolarT04}.
The surface gradient $\vec{p}$ is related to the tilt angle $\theta$ as $\vec{p}=\pm \tan \theta$.
It should be noted that the Andreev's free energy is similar to the equilibrium crystal shape (ECS), which is the shape with the least total surface free energy.
The ECS is obtained by the Landau-Andreev method \cite{andreev, landau}.
Alternatively, the ECS is obtained from the surface tension with the Wulff construction \cite{laue, bcf, herring, mackenzie} based on the Wulff theorem.

In Fig.~\ref{gampolarT04}, there are values for the (001) and (111) surfaces \cite{akutsu11JPCM}.
The vicinal surfaces between the (001) surface and the (111) surface did not appear, because these vicinal surfaces are thermodynamically unstable. 
This is the characteristic of the profile of the faceted macrostep in the step-faceting zone at equilibrium \cite{akutsu16}.

%\section{Monte Carlo results}

\section{Monte Carlo Results \label{MCresults}}
%Monte Carlo method

\subsection{Monte Carlo method}

To study the non-equilibrium steady-state with macrosteps, the vicinal surface of the following Hamiltonian with a fixed number $N_{\rm step}$ of steps was investigated using the Monte Carlo method with the Metropolis algorithm: 
\beq
{\mathcal H}_{\rm non eq}= {\mathcal H}_{\rm p-RSOS} 
- \Delta \mu \sum_{n,m} [h(n,m,t+1)-h(n,m,t)], \label{hamil_neq}
\eeq
where $t$ is the time measured by the Monte Carlo steps per site (MCS/site).
When $\Delta \mu>0$, the crystal grows, whereas when $\Delta \mu<0$, the crystal recedes (evaporates, dissociates, or melts).

The explicit procedure for application of the Monte Carlo method in this study is as follows.
At the initial time, the vicinal surface is set with an initial configuration.
The lattice site to be updated is then randomly selected. The surface structure is updated non-conservatively using the Monte Carlo method with the Metropolis algorithm. 
With the RSOS restriction taken into consideration, the structure is updated with probability 1 when $\Delta E \leq 0$ and with probability $\exp(- \Delta E /\kBT )$ when $\Delta E > 0$, where $\Delta E= E_{f}-E_{i}$, $E_{i}$ is the energy of the present configuration, and $E_{f}$ is the energy of the updated configuration. The energy is calculated using the Hamiltonian (Eq.~(\ref{hamil_neq})).

A periodic boundary condition was imposed in the direction parallel to the steps. 
In the direction normal to the steps, the lowest side of the structure was connected to the uppermost side by the addition of a height with a number $N_{\rm step}$ of steps. 

\begin{figure*}%[h]%[htbp]
%\begin{center}
\centering
\includegraphics[width=15.0 cm,clip]{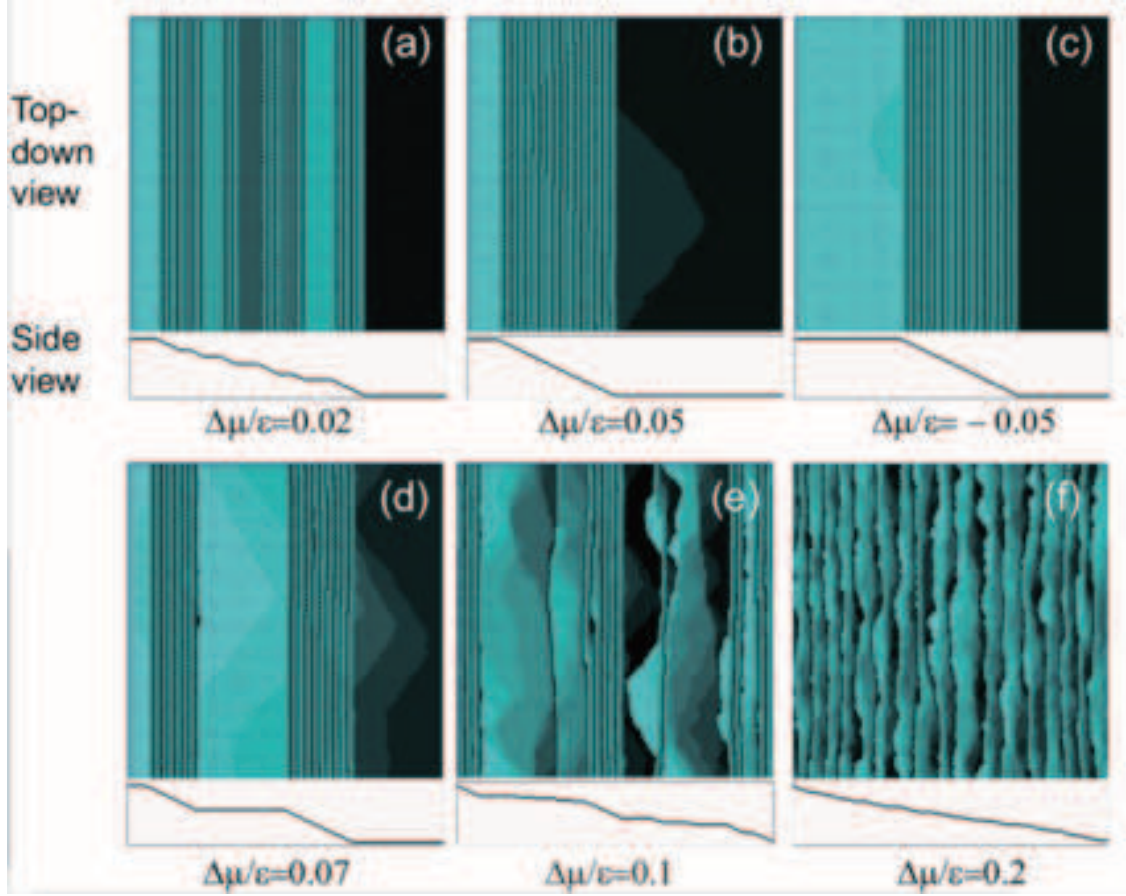}%
%\end{center}
\caption{
(Color online)
Snapshots of the surface from Monte Carlo simulation of a vicinal surface at $4 \times 10^8$ Monte Carlo MCS/site. The surfaces are inclined towards the $\langle 111 \rangle$ direction.
The initial configuration is a train of elementary steps (TS) with equal distance.
(a) $|\Delta \mu| <\Delta \mu_f(\tau, L)$.
(b) and (c) $\Delta \mu_f(\tau, L)<|\Delta \mu| \leq \Delta \mu_{co}(L)$.
(d) and (e) $\Delta \mu_{co}(L)<|\Delta \mu| <\Delta \mu_R(L)$.
(f) $\Delta \mu_R(L) < |\Delta \mu|$.
$\kBT/\epsilon=0.4$. The number of steps ($N_{\rm step}$) equals 180.
Size: $240 \sqrt{2} \times 240 \sqrt{2}$.
$\epsilon_{\rm int}/\epsilon=-0.9$.
The surface height is represented by brightness with 10 gradations, where brighter regions are higher.
The darkest areas next to the brightest areas represent terraces that are higher by a value of unity because of the finite gradation. 
The lines of side view are drawn with respect to the height along the bottom lines in the top-down view.
}  
\label{surfdatT04}
\end{figure*}

Two types of initial configuration for the steps were prepared: a train of elementary steps with equal distance (the TS configuration) and one macrostep with the (111) side surface (the MS configuration).
In both configurations, the mean surface slope $\bar{p}=N_{\rm step}/L$ is kept constant, where $L$ is the linear size of the system.
Figure \ref{surfdatT04} shows snapshots of the vicinal surfaces at $4 \times 10^8$ Monte Carlo MCS/site, where the initial configuration is TS.

%%%%% FIGURE  %%%%%

%%%%% FIGURE  %%%%%

%%%%% FIGURE  %%%%%

\begin{figure*}%[h]%[htbp]
%\begin{center}
\centering
\includegraphics[width=15.0 cm,clip]{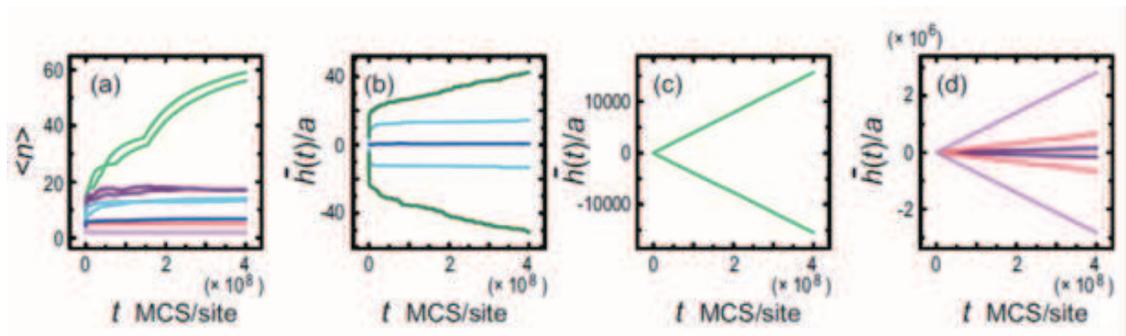}%
%\end{center}
\caption{
Time evolution of $\langle n \rangle$ and $\bar{h}(t)$.
The initial configuration is the TS configuration.
(a) Time evolution of $\langle n \rangle$.
(b--d) Time evolution of $\bar{h}(t)$.
Dark blue lines: $\Delta \mu=0$.
Light blue lines: $\Delta \mu /\epsilon= \pm0.02$.
Green lines: $\Delta \mu /\epsilon= \pm0.03$.
Light green lines $\Delta \mu /\epsilon= \pm0.05$.
Dark purple lines: $\Delta \mu /\epsilon= \pm0.07$.
Light purple lines: $\Delta \mu /\epsilon= \pm0.1$.
Pink lines: $\Delta \mu /\epsilon= \pm0.2$.
Size: $240 \sqrt{2} \times 240 \sqrt{2}$.
$N_{\rm step}=180$.
$\epsilon_{\rm int}/\epsilon=-0.9$.
}  
\label{t_depT04}
\end{figure*}
%%%%% FIGURE  %%%%%

\subsection{Time evolution of surface height}

%%%%% FIGURE  %%%%%

\begin{figure*}%[h]%[htbp]
%\begin{center}
\centering
\includegraphics[width=14.0 cm,clip]{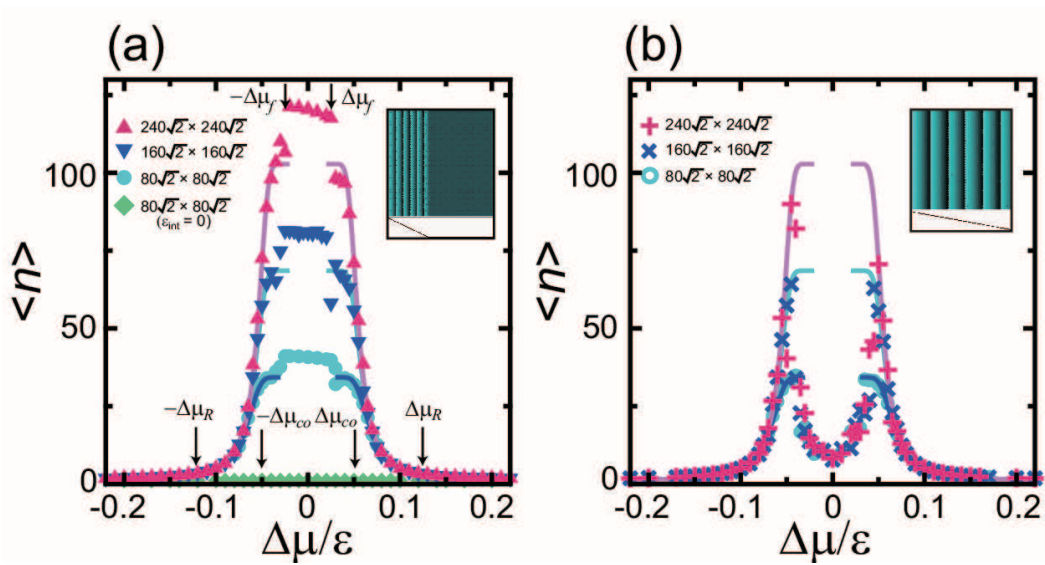}%
%\end{center}
\caption{
(Color online)
Average height of merged steps $\langle n \rangle$.
(a) Initial configuration with a macrostep (MS).
(b) Initial configuration with a train of steps (TS) in equal distance.
The inset shows the top view and the side view of the initial configuration of the vicinal surface.
Mean surface slope: $\bar{p}=0.530$.
$\kBT/\epsilon=0.4$.
$\epsilon_{\rm int}/\epsilon=-0.9$.
The initial $2 \times 10^8$ MCS/site were discarded.
The values are averaged over the following $2 \times 10^8$ MCS/site.
Solid lines: Eqs.~(\ref{eq_nav2}) and (\ref{eq_1/z}) with $N_{\rm step}=180$, 120, and 60 from top to bottom.
$\Delta \mu_{f}(\tau,L)$, $\Delta \mu_{co}(L)$, and $\Delta \mu_R(L)$ are listed in Table \ref{table_def}, where $\tau=4\times 10^8$ is the observation time, and $L=240\sqrt{2}$ is the linear size of the system.
}  
\label{nav}
\end{figure*}
%%%%% FIGURE  %%%%%

%%%%% FIGURE  %%%%%

\begin{figure}%[h]%[htbp]
%\begin{center}
\centering
\includegraphics[width=7.0 cm,clip]{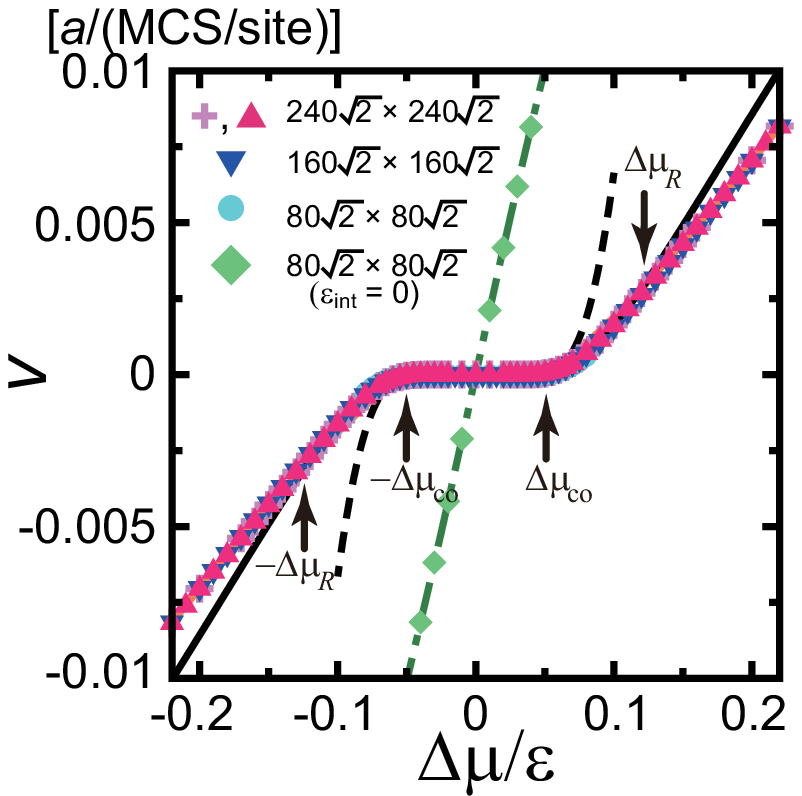}%
%\end{center}
\caption{
(Color online)
Growth rate of the surface $V$.
The initial configuration is MS.
$\kBT/\epsilon=0.4$.
$\epsilon_{\rm int}/\epsilon=-0.9$.
Crosses show the Monte Carlo results obtained with the TS initial configuration.
$\Delta \mu_{co}$ and $\Delta \mu_R(L)$ are in Table \ref{table_def}.
Dashed lines: Eq.~(\ref{eq_V_2D2}).
Solid lines: Eq.~(\ref{eq_Vsurf}).
Pale solid lines: Eq.~(\ref{eq_V_power}).
Dash-dotted line: Eq.~(\ref{eq_vstep4}) with $p_1=0.483$.
}  
\label{GR}
\end{figure}
%%%%% FIGURE  %%%%%

To study the characteristics of the vicinal surface at the mesoscopic scale (20 nm to 500 nm), the average height of merged steps \cite{akutsu11JPCM} and the growth rate of the surface were calculated using the Monte Carlo method.
To evaluate the size of a macrostep in detail, we consider the number of elementary steps in a locally merged step $n$.
The average height of the merged steps is obtained as follows: 
\beq
\langle n \rangle =\sum_{\tilde{y}}  \sum_{\tilde{x}}|n_{\tilde{x}}(\tilde{y})|/  [ \sum_{\tilde{y}} n_{\rm step}(\tilde{y}) ]\approx N_{\rm step}/{\langle n_{\rm step} \rangle },
\eeq
where $\tilde{x}$ is selected as the $\langle 110 \rangle$ direction (normal to the mean step-running direction), $\tilde{y}$ as the $\langle \bar{1}10 \rangle$ direction (along the mean step-running direction), $N_{\rm step}$ is the total number of elementary steps, and $n_{\rm step}$ is the number of merged steps.
Time evolutions of $\langle n \rangle$ are shown in Fig.~\ref{t_depT04}(a).
After $2 \times 10^8$ MCS/site, $\langle n \rangle$ is almost constant for $\Delta \mu/\epsilon>0.05$.
Therefore, the values obtained within 0--$2 \times 10^8$ MCS/site were discarded, and $n$ was averaged over successive $2 \times 10^8$ MCS/site.
Figure \ref{nav} shows the $\Delta \mu$ dependence of $\langle n \rangle$.

To estimate the growth rate of the surface $V$, the average surface height $\bar{h}(t)$, was calculated, where 
\beq
\bar{h}(t)= (1/ {\mathcal N})\sum_{n,m}h(n,m).
\eeq
Time evolutions of $\bar{h}(t)$ are also shown in Fig.~\ref{t_depT04}(b)--(d). 
$\bar{h}(t)$ increases or decreases linearly as $t$ increases for $0.035 \simless |\Delta \mu|/\epsilon$. 
At $0.025 \simless |\Delta \mu|/\epsilon \simless 0.035$, the surface height increases intermittently.
For $|\Delta \mu/\epsilon| \simless 0.025$, the surface does not grow/recede.

The growth rate of the surface $V$, is defined as:
\beq
V=[\bar{h}(t_{\rm max})-\bar{h}(t_0)]/(t_{\rm max}-t_0),
\eeq
where $t_0$ and $t_{\rm max}$ are $2 \times 10^8$ MCS/site and $4 \times 10^8$ MCS/site, respectively.
Figure \ref{GR} shows the $\Delta \mu$ dependence of $V$.

\begin{table}
\caption{\label{table_def} Characteristic driving forces.}
\begin{ruledtabular}
\begin{tabular}{clc}
Symbol& value$/\epsilon$ & $L/(\sqrt{2}a)\footnote{$L$ is the linear size of the system. $a=1$.}$ \\
\hline
\vspace{-0.7em}&&\\
$\Delta \mu_y(L) $\footnote{Yield point of the self-detachment of steps from a macrostep. (\S \ref{sec_yielding})} & $0.018\pm0.006$ &240    \\
&$0.020\pm0.006$ &160  \\
&$0.023 \pm0.004$& 80  \\
$\Delta \mu_{f}(\tau, L)$\footnote{Freezing point of step-detachments. $\tau=4\times 10^8 $ MCS/site is the observation time.  (\S \ref{sec_2Dnucleation}) } & $0.023\pm0.007$ &240 \\ 
 & $0.025\pm0.007$ &160  \\ 
 & $0.027\pm0.007$ &80\\ 
$\Delta \mu_{co}(L)$\footnote{Crossover point from 2D nucleation mode to successive step-detachment mode. (\S \ref{k-roughening}) } & $0.050\pm0.007$ &240 \\ 
&$0.051\pm0.007$ &160 \\
&$0.054\pm0.007$&80 \\
$\Delta \mu_R(L)$\footnote{Crossover point between the step-detachment mode and kinetically roughened mode. (\S \ref{sec_yielding}) }$-\Delta \mu_{co}(L)$ & $0.071\pm0.005$ & -- \\
\end{tabular}
\end{ruledtabular}
\end{table}

\subsection{Characteristic driving forces}

There are several characteristic driving forces in the non-equilibrium steady-state, which are listed in Table \ref{table_def}.
Each value will be explained in detail in the following sections.
Here, we will explain them briefly. 

First, $\Delta \mu_R(L)$ is defined as the minimum $|\Delta \mu|$ so that the macrostep disappears (Fig.~\ref{surfdatT04})(f), \S \ref{sec_yielding}).
$L$ is the linear size of the system.
In this region, the surface roughens kinetically.
The velocity of the surface $V$, and the average height of the merged step $\langle n \rangle$, exhibits power law behavior with respect to $|\Delta \mu|-\Delta \mu_{co}(L)$ if we introduce $\Delta \mu_{co}(L)$ (\S \ref{k-roughening}).
$\Delta \mu_{co}(L)$ is determined so that the log-log plot with respect to $V$ {\it vs.} $|\Delta \mu|-\Delta \mu_{co}(L)$ becomes linear. 
The Monte Carlo results show that $\Delta \mu_R(L)-\Delta \mu_{co}(L)$ is not significantly dependent on $L$ (Table \ref{table_def}).

For $\Delta \mu_{co}(L)< |\Delta \mu|< \Delta \mu_R(L)$, elementary steps detach from macrosteps and successively attach to the macrostep (Figs.~\ref{surfdatT04}(d) and (e), \S \ref{sec_detachment}).

Next, we define $\Delta \mu_{f}(\tau, L)$, where $\tau$ is the observation time.
$\Delta \mu_{f}(\tau, L)$ is defined as the maximum $|\Delta \mu|$ so that growth/recession is inhibited during the observation time. 
In the region $|\Delta \mu| < \Delta \mu_{f}(\tau, L)$, the growth/recession of the vicinal surface freezes due to the finite size and time effect (\S \ref{sec_2Dnucleation}).
The surface does not reach the true non-equilibrium steady-state; therefore, the surface morphology is dependent on the initial configuration of the surface.
$\langle n \rangle$ in Figs.~\ref{nav}(a) and (b) for $|\Delta \mu| < \Delta \mu_{f}(\tau, L)$ is strongly dependent on the initial configuration.

For $\Delta \mu_{f}(\tau, L)< |\Delta \mu|< \Delta \mu_{co}(L)$, the surface grows/recedes intermittently in the manner of 2D heterogeneous nucleation (\S \ref{sec_2Dnucleation}, Figs.~\ref{surfdatT04}(b) and (c)).
 $\langle n \rangle$ of the TS initial condition is smaller than $\langle n \rangle$ of the MS initial configuration.
On the other hand, the growth rates with the TS initial configuration agree well with those for the MS initial configuration. 
An island on the (001) surface or an island on the (111) surface is formed at the edge of the faceted macrostep, as shown in Figs.~\ref{surfdatT04}(b) and (c).

Finally, we define $\Delta \mu_{y}(L)$ (\S \ref{sec_yielding}).
$\Delta \mu_{y}(L)$ is defined as the minimum value of $|\Delta \mu|$ so that steps are spontaneously and successively detached from a macrostep.
$\Delta \mu_{y}(L)$ is determined by the extrapolation of $p_1$, which is the surface slope contacted with the (111) faceted surface.
For $|\Delta \mu| < \Delta \mu_{y}(L)$, the (001) surface ($p_1=0$) contacts with the (111) surface as a ``terrace'', whereas for $|\Delta \mu| > \Delta \mu_{y}(L)$, the surface with a slope of $p_1 \neq 0$ contacts with the (111) surface as a ``terrace''.

\section{Step detachment \label{analysis}}

\subsection{2D heterogeneous nucleation \label{sec_2Dnucleation}}

\subsubsection{Growth/recession rate of the surface \label{sec_2DV}}

%%%%% FIGURE  %%%%%

\begin{figure}%[h]%[htbp]
%\begin{center}
\centering
\includegraphics[width=8 cm,clip]{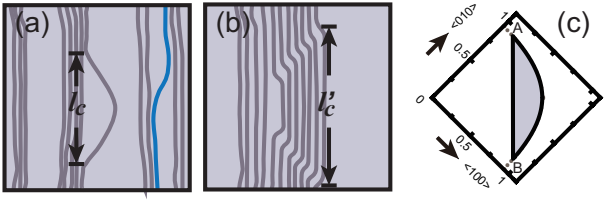}%
%\end{center}
\caption{
(Color online)
Schematic figure of 2D critical nuclei under growth conditions.
(a) A critical nucleus on the (001) surface. 
$l_c$ is the linear size of the critical nucleus.
(b) A critical nucleus on the (111) surface.
$l_c'$ is the linear size of the critical nucleus.
Gray lines: Elementary steps on a surface.
Light blue line: A step containing a ``bridge'' which runs across the terrace.
During growth, the 2D nuclei appear at the lower section line of the (001) and (111) surfaces of a macrostep, whereas during recession, the 2D nuclei appear at the upper section line of the (001) surface and at the (111) surface side of a macrostep.
}  
\label{nuclei}
\end{figure}
%%%%% FIGURE  %%%%%

For $\Delta \mu_f(\tau,L) < |\Delta \mu| < \Delta \mu_{co}(L)$ (Table \ref{table_def}), the surface grows/recedes by step-detachment through 2D ``heterogeneous'' nucleation (Fig.~\ref{nuclei}).
This means that the growth/recession of the surface occurs intermittently.
 The nuclei were created at the lower/upper side of the macrostep-edge (Figs.~\ref{surfdatT04}(b) and (c)) in the growth/recession mode, respectively.
We describe the side length of the critical nucleus on the (001) surface as $l_c$, and that on the (111) surface as $l_c'$ (Fig.~\ref{nuclei}).
Figure \ref{nuclei} shows $l_c \sim l_c'$ of the 2D critical nucleus for the step-detachment with the growth of the surface ($\Delta \mu >0$).
For $\Delta \mu <0$, an elementary step detaches from the upper side of the (111) surface by forming a critical {\it negative} nucleus.
The shape of the critical negative nucleus is similar but reversed to the shape shown in Fig.~\ref{nuclei}(c) (also see Fig.~\ref{surfdatT04}(c)).

The Gibbs free energy $G(l)$ (or $G(l')$) of the island attached to the faceted macrostep is expressed as:
\beq
G(l) = -|\Delta \mu| S(l) +\Gamma(l) - l \gamma_2^{(110)} \label{eq_gibbs},
\eeq
where $S(l)$ is the area of the island, $\Gamma(l)$ is the total step free energy at the edge of the island of the elementary step, and $\gamma_2^{(110)}$ is the step free energy of the doubly merged step.
For the critical nucleus, the Gibbs free energy is a minimum with respect to the shape, but is a maximum with respect to the size \cite{becker}.
The shape of the critical nucleus is similar to the equilibrium island shape (Fig.~\ref{nuclei}(c)).
The size of the critical nucleus is determined as explained in Appendix \ref{2Dcritical},
i.e.,
\beq
l_c=l_{c,0}\epsilon/|\Delta \mu|, \quad G(l_c)=G(l_{c,0})\epsilon/\Delta \mu, \label{eq_size2D} 
\eeq 
where $l_{c,0}$ and $G(l_{c,0})$ are $l_{c}$ and $G(l_{c})$ for $|\Delta \mu|/\epsilon=1$ obtained by Eqs.~(\ref{eq_Gamma0}) and (\ref{eq_G0}) as:
\beq
G(l_{c,0})=(\Gamma(l_{c,0}) - l_{c,0} \gamma_2^{(110)})/2 = \epsilon S(l_{c,0}).
\label{eq_Glc0}
\eeq

%%%%% FIGURE  %%%%%

\begin{figure}%[h]%[htbp]
%\begin{center}
\centering
\includegraphics[width=6.5 cm,clip]{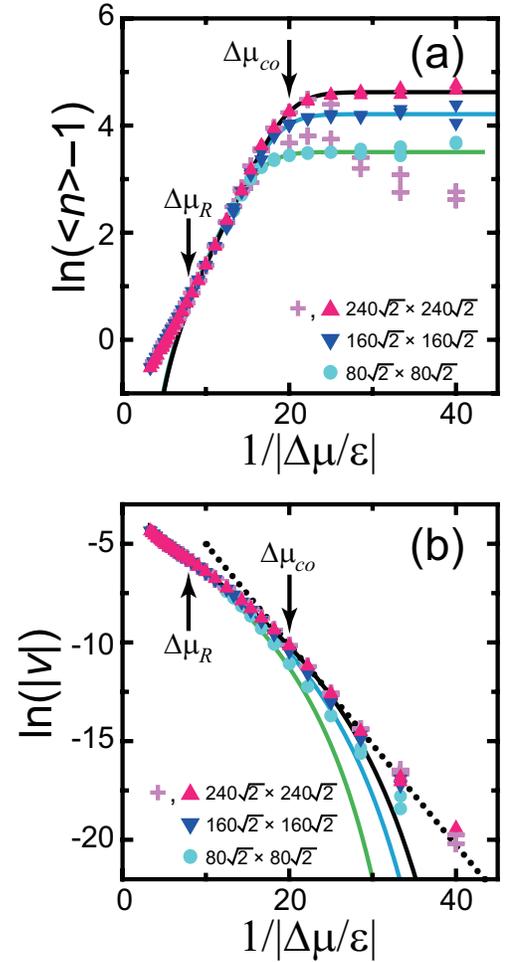}%
%\end{center}
\caption{
(Color online)
(a) Semi-logarithmic plot of $\langle n \rangle -1$.
Solid lines: Eqs.~(\ref{eq_p1_black}) and (\ref{eq_nav2}) with $N_{\rm step}= 180$, 120, and 60 from the top to the bottom.
(b) Semi-logarithmic plot of the absolute value of the surface growth rate.
Dotted line: Eq.~(\ref{eq_V_2D2}).
Solid lines: $V =p_1 v_{\rm step} $ with Eqs.~(\ref{eq_p1_black}) and (\ref{eq_vstep2}), where $N_{\rm step}= 180$, 120, and 60 from the top to the bottom.
Crosses show the Monte Carlo results obtained with the TS initial configuration.
$\kBT/\epsilon=0.4$.
$\epsilon_{\rm int}/\epsilon=-0.9$.
Values were averaged over $2 \times 10^8$ MCS/site after first discarding $2 \times 10^8$ MCS/site.
}  
\label{lnyT04}
\end{figure}
%%%%% FIGURE  %%%%%

\begin{table}
\caption{\label{table_tn} Characteristic lengths and times.
}
\begin{ruledtabular}
\begin{tabular}{lcccc}
$\Delta \mu/\epsilon$ & $l_c/a$ & $t_n$ [MCS/site]\footnote{$L=240\sqrt{2}a$. $a=1$.} & $l_d/a$\footnote{$v_t$ is assumed to be given by Eq.~(\ref{eq_vstep4}). } & $t_d$ [MCS/site]  \\
&Eqs.~(\ref{eq_size2D})  & Eq.~(\ref{eq_tn})   & Eq.~(\ref{eq_ld}) & Eq.~(\ref{eq_td})\\
& \& (\ref{eq_lc0}) &  \& (\ref{eq_V_2D2})  &  & \\
\hline
\vspace{-0.5em}&&&\\
0.01 & 106 &  $8.3\times10^{21}$ & $3.2\times10^{9}$ & $\underline{1.7\times10^{12}}$\\
0.02 & 53 &  $\underline{7.4\times10^{10}}$  & $1.2\times10^4$ & $3.1\times10^{7}$\\
0.03 & 35 &  $1.4\times10^{7}$ &  $4.4\times10^3$ & $7.6\times10^{5}$\\
0.04 & 27 &  $2.2\times10^{5}$  &  $884$ & $1.1\times10^{5}$\\
0.05 & 21 &  $1.7\times10^{4}$  &  $\underline{345}$ & $3.5\times10^{4}$\\
0.06 & 18 &  $3.2\times10^{3}$  &  $188$ & $1.6\times10^{4}$\\
0.1  & 11 &  $106$  &  $60$ & $3.0\times10^{3}$\\
0.15 & 7.1 &  $19$  &  $37$ & $1.2\times10^{3}$\\
%0.2 & 5.4 &  $8.3$  & $22$ & $1.3\times10^{2}$
\end{tabular}
\end{ruledtabular}
\end{table}

From the classical nucleation theory \cite{becker,ookawa,pimpinelli}, the nucleation frequency $I_n$ is expressed as follows:
\beq
I_n=ZN_0/C \exp[- G(l_c)/\kBT],
\eeq  
where $Z$ is the Zeldovich factor, $N_0=1/(\sqrt{2}a)$ is a lattice-point density, and $C$ is a coefficient relating to geometry.
The waiting time for a single nucleation $t_n$ is 
\beq
t_n=1/(I_n L) %\nonumber\\
= \frac{\sqrt{2}C}{ZL} \exp[g^*/\Delta \mu],
\label{eq_tn}
\eeq
where $g^*/\Delta \mu=  G(l_c)/(\kBT) $.
In the limit of $L \rightarrow \infty$, $t_n$ reduces to zero because 2D nuclei are formed somewhere at the step edge of the macrostep.
The growth rate $V$ is expressed using $t_n$ as 
\beq
|V|= a/t_n= \frac{L Z}{\sqrt{2}C} \exp[-g^*/\Delta \mu]. \label{eq_V_2D}
\eeq

$\ln(|V|)$ is shown with the horizontal axis being $1/(|\Delta \mu|/\epsilon)$ in Fig.~\ref{lnyT04}(b).
The Monte Carlo results gave
\beqa
|V|= 1.07 \exp[-0.509/(|\Delta \mu|/\epsilon)], \label{eq_V_2D2}
\eeqa
by fitting a line in Fig.~\ref{lnyT04}(b) to the values of $\Delta \mu_y(L)< |\Delta \mu|< \Delta \mu_{co}(L)$ with $L=240 \sqrt{2}$.
Therefore, we have $g^*_{\rm MC}=0.509\epsilon$.
This line is shown by the dotted line in Fig.~\ref{lnyT04}(b).
$Z/(\sqrt{2}C)=4.5 \times 10^{-3}$ was also obtained from the prefactor on the right-hand side of Eq.~(\ref{eq_V_2D}).
Using this value with Eq.~(\ref{eq_tn}), we show the explicit values for $t_n$ with $L=240 \sqrt{2}$ for several driving forces (Table \ref{table_tn}).

To numerically obtain $l_c$, $l_{c,0}$ and $G(l_{c,0})$ were calculated using the 2D Ising model \cite{akutsu86,akutsu90,akutsu92} at $\kBT/\epsilon=0.4$.
The results were $\Gamma(l_{c,0})=\int_A^B \gamma(\theta) {\rm d}l \approx 1.165 \epsilon$ and $\gamma_2^{(110)}=\sqrt{2}(\epsilon+\epsilon_{\rm int}/2)\approx 0.7778\epsilon/a$, where $\theta$ is the mean tilt angle of an elementary step relative to the $\langle 010 \rangle$ direction.
From Eq.~(\ref{eq_Glc0}), we obtain 
\beq
l_{c,0} \approx 1.0649a, \quad G(l_{c,0})=0.1686\epsilon.\label{eq_lc0}
\eeq
$g^*$ obtained from Eq.~(\ref{eq_lc0}) is 
\beq
g^*_{\rm Ising}=G(l_{c,0})|\Delta \mu|/\kBT=0.4218\epsilon. \label{eq_g*}
\eeq
Table \ref{table_tn} shows the size of the $l_c$ for several $\Delta \mu$.

$g^*_{\rm Ising}$ is slightly smaller than $g^*_{\rm MC}$ obtained by the Monte Carlo method.
The reason for this seems to be a reduction of the entropy due to finite size of the critical island.
The values of $l_c$ in Table \ref{table_tn} indicate the size of the critical island to be less than approximately 100 in the present simulations.
With such a short length, the entropy term in the step tension becomes smaller than the entropy term in the step tension with infinite length.
Therefore, $g^*_{\rm MC}$ becomes larger than $g^*_{\rm Ising}$ calculated for the infinite length of the domain boundary line.

Due to the long waiting time for the 2D heterogeneous nucleation at the edge of the faceted macrostep, the surface growth occurs intermittently.
The intermittent growth/recession can be observed explicitly in the case of $|\Delta \mu/\epsilon|=0.03$, as shown in Fig.~\ref{t_depT04}(c).

From $t_n$ in Table \ref{table_tn}, the waiting time exceeds $4 \times 10^8$ MCS/site for $|\Delta \mu| \simless \Delta \mu_f(\tau, L)$.
Therefore, for $|\Delta \mu| <\Delta \mu_f(\tau, L)$, the surface cannot grow/recede due to this finite time effect.

\subsubsection{Average height of merged steps}

To understand the $\Delta \mu$ dependence of the vicinal surface morphology we consider a step--attachment--detachment model for the time evolution of $\langle n \rangle$ \cite{akutsu17}:
\beq
\frac{\partial \langle n \rangle}{\partial t}= n_+ - n_-,
\eeq
where $n_+$ is the rate when the elementary steps catch up to a macrostep, and $n_-$ is the rate when the elementary steps detach from a macrostep.
When $n_+<n_-$, a macrostep dissociates, 
whereas when $n_+>n_-$, $\langle n \rangle$ increases up to $N_{\rm step}$, where $N_{\rm step}$ is the total number of elementary steps on the surface.
In this case, $n_-$ limits the growth/recession rate of the surface.
At steady-state, $n_+ = n_- = V/a$, where $a$ is the height of the elementary step.

In the region of $\Delta \mu_{f}(\tau, L)<|\Delta \mu|<\Delta \mu_{co}(L)$, $n_+$ is considered to be $n_+ \approx \rho_1 v_1$, where $\rho_1$ is the density of elementary steps on the ``terrace'', and $v_1$ is the step velocity of an elementary step perpendicular to the mean running direction of the step.
On the other hand, $n_-$ is proportional to the 2D heterogeneous nucleation rate.
Therefore, $n_+> n_-$ is expected because the growth rate of an elementary step $v_1$ is relatively large (for example, \S \ref{sec_stepV}, Fig.~\ref{VstepT04}).  
Therefore, after a sufficiently long time, elementary steps merge to form a single macrostep (Figs.~\ref{surfdatT04}(b) and (c)).

It is noted that the morphology of the surface also freezes for $|\Delta \mu| < \Delta \mu_{f}(\tau, L)$.
$\langle n \rangle$ is strongly dependent on the initial configuration of the surface (Fig.~\ref{surfdatT04}(a), Figs.~\ref{nav}(a) and (b)).

\subsection{Successive step-detachment \label{sec_detachment}}

\subsubsection{Multi-nucleations \label{sec_multi}}

From Fig.~\ref{lnyT04}(b) of the growth/recession rate of the surface for $\Delta \mu_{co}<|\Delta \mu|<\Delta \mu_R(L)$, the slope of the line changes at a crossover point, $\Delta \mu_{co}$.
In addition, $\langle n \rangle$ is no longer constant but decreases as $1/|\Delta \mu|$ decreases.
In this subsection, we will apply the classical multiple heterogeneous nucleation theory to the Monte Carlo results.

%%%%% FIGURE  %%%%%

\begin{figure}%[h]%[htbp]
%\begin{center}
\centering
\includegraphics[width=3 cm,clip]{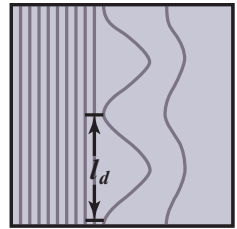}%
%\end{center}
\caption{Schematic figure of multi-nucleation under growth conditions.
}  
\label{multi_nuclei}
\end{figure}
%%%%% FIGURE  %%%%%

We assume some critical nuclei arise at the edge of a macrostep with a mean equal distance $l_d$ (Fig.~\ref{multi_nuclei}).
The step-detachment time $t_d$, is then approximated as
\beq
t_d=l_d/(2 v_t)=1/(I_n l_d), \label{td_def}
\eeq
where $v_t$ is the step unzipping velocity.
From Eq.~(\ref{td_def}), $l_d$ is expressed using $v_t$ and $I_n$ as
\beq
l_d= \sqrt{\frac{2 v_t}{I_n}}% \nonumber \\
= \sqrt{\frac{2\sqrt{2} v_tC}{Z}} \exp[g^*/(2\Delta \mu)]. \label{eq_ld} 
\eeq
By substituting Eq.~(\ref{eq_ld}) into Eq.~(\ref{td_def}), we obtain
\beq
t_d=1/\sqrt{2 v_t I_n}= \sqrt{\frac{C}{\sqrt{2} v_tZ}} \exp[g^*/(2\Delta \mu)].  \label{eq_td}
\eeq
The explicit values of $l_d$ and $t_d$ are calculated using Eqs.~(\ref{eq_ld}) and (\ref{eq_td}) with $g^*$ being $g^*_{\rm Ising}$ and $Z/(\sqrt{2}C)$ being $4.5 \times 10^{-3}$, where $v_t$ is assumed to equal $v_{\rm step, RSOS}(|\Delta \mu|)$ (\S \ref{sec_stepV}, Eq.~(\ref{eq_vstep4})), which are given in Table \ref{table_tn}.

The growth/recession rate of the surface is then obtained by
\beqa
|V|&=& a/t_d=a\sqrt{2 v_t I_n}, \nonumber \\
&=& \sqrt{\frac{\sqrt{2} v_tZ}{C}} \exp[- g^*/(2\Delta \mu)]. \label{eq_Vmulti}
\eeqa
From $v_t \propto \Delta \mu$, we expect
\beq
|V| \propto \sqrt{\Delta \mu} \exp[-g^*/(2\Delta \mu)] \label{eq_Vmulti2}
\eeq
 approximately for $\Delta \mu_{co}(L)< |\Delta \mu| < \Delta \mu_R(L)$.
In Fig.~\ref{lnyT04}(b), the slope of the Monte Carlo results in this region seems to be smaller than the slope for $|\Delta \mu| < \Delta \mu_{co}(L)$.
However, the Monte Carlo results for $|V|$ bend around $\Delta \mu_{co}< |\Delta \mu| < \Delta \mu_R(L)$.

It is interesting that $t_d \simless t_n$ for $\Delta \mu_{co}(L)<|\Delta \mu|$.
The 2D nuclei are randomly formed relatively often at the edge of the macrostep.
The elementary step formed by 2D nucleation advances/recedes by detachment from the macrostep.
After a certain time, the elementary step may be pulled back to the facet edge with some probability.
Therefore, for $\Delta \mu_{co}(L)< |\Delta \mu| < \Delta \mu_R(L)$, the detachment of an elementary step from the edge of the macrostep limits the growth/recession rate of the vicinal surface.

In this manner, classical 2D heterogeneous multi-nucleation is found to explain the phenomena roughly.
However, with a slight modification, agreements between the Monte Carlo results and the expressions based on classical 2D heterogeneous multi-nucleation with respect to $|V|$ and $\langle n \langle$ are significantly improved, 
as demonstrated in the following sub-subsections.

%%%%% FIGURE  %%%%%

\begin{figure}%[h]%[htbp]
%\begin{center}
\centering
\includegraphics[width=7.0 cm,clip]{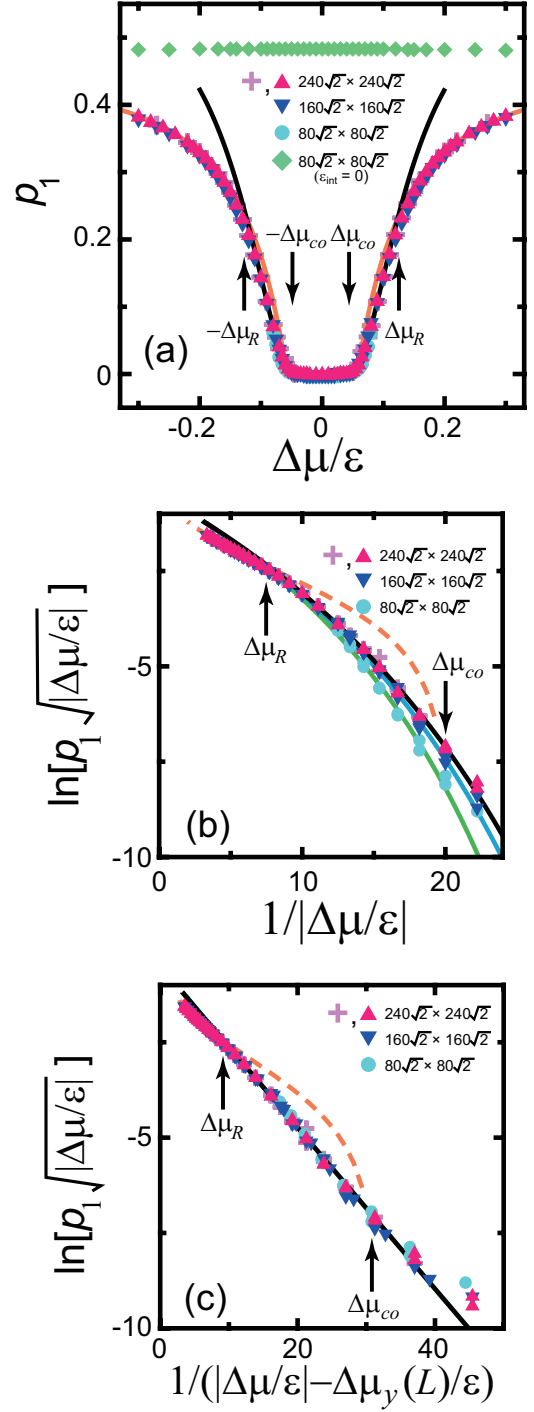}%
%\end{center}
\caption{
(Color online)
(a) $\Delta \mu$ dependence of the surface slope $p_1$, of the ``terrace''. 
(b) $\ln[p_1 \sqrt{|\Delta \mu/\epsilon|}]$ {\it vs.} $1/|\Delta \mu/\epsilon|$.
(c)$\ln[p_1 \sqrt{|\Delta \mu/\epsilon|}]$ {\it vs.} $1/|\Delta \mu/\epsilon|-\Delta \mu_y(L)/\epsilon$.
$p_1$ for the Monte Carlo data is calculated using Eq.~(\ref{eq_p1}) with $\langle n \rangle$.
Crosses indicate the Monte Carlo results obtained with the TS initial configuration.
Dark solid lines: Eq.~(\ref{eq_p1_black}) with $L=240\sqrt{2}$.
Pale solid lines in (a) and dashed lines in (b) and (c): $p_1= V/v_{\rm step}$ with Eqs.~(\ref{eq_vstep3}) and (\ref{eq_V_power}).
Pale solid lines in (b): Eq.~(\ref{eq_p1_black}) with $L=160\sqrt{2}$ (upper) and $L=80\sqrt{2}$ (lower).
$\kBT/\epsilon=0.4$.
$\epsilon_{\rm int}/\epsilon=-0.9$.
}  
\label{p1T04}
\end{figure}
%%%%% FIGURE  %%%%%

\subsubsection{Surface slope of the ``terrace''\label{sec_yielding}}

Let us call the surface that contacts the faceted macrostep the ``terrace''.
At equilibrium in the step-faceting zone, the ``terrace'' is the (001) surface, which characterizes the profile of the faceted macrostep \cite{akutsu16-3}.
In the case of a non-equilibrium steady-state, the change of the ``terrace'' slope changes the dynamics of the vicinal surface.
In this sub-subsection, we explain how the surface slope $p_1$, is connected to the size of the merged step $\langle n \rangle$, the yielding point $\Delta \mu_y(L)$, and the crossover point to a kinetically roughened surface, $\Delta \mu_R(L)$.

For every $t_d$, an elementary step is detached from the macrostep at the edge of the macrostep (Fig.~\ref{multi_nuclei}).
The detached elementary steps form a vicinal surface with the slope $p_1$, which contacts the macrostep.
After several calculations (Appendix \ref{sec_p1-<n>}), the surface slope is described by $\langle n \rangle$, as follows: \cite{akutsu17}:
\beqa
 p_1&=& \sqrt{2} /\left( \frac{\sqrt{2}-\bar{p}}{\bar{p}z} +1\right),\nonumber \\ 
%\quad 
z&=& \frac{1}{\langle n \rangle}-\frac{N_m}{N_{\rm step}}, \label{eq_p1}
\eeqa
where $N_m$ is the number of macrosteps in the simulated system.

Using Eq.~(\ref{eq_p1}) with the assumption $N_m=1.75$, $p_1$ is calculated from $\langle n \rangle$ obtained from the Monte Carlo calculation (Fig.~\ref{p1T04}(a)).
Keeping Eq.~(\ref{eq_Vmulti2}) in mind, we present $\ln [p_1 \sqrt{|\Delta \mu/\epsilon|}]$ {\it vs.} $1/|\Delta \mu/\epsilon|$ (Fig.~\ref{p1T04}(b)).
Here, the Monte Carlo results are not straight lines.
In addition, for small $|\Delta \mu|$, the Monte Carlo results reveal the size dependence.

Here, let us introduce $\Delta \mu_y(L)$ so that the Monte Carlo results are well reproduced by a straight line (Fig.~\ref{p1T04}(c)).
For $\mu_{co}(L)<|\Delta \mu|< \mu_R(L)$, the best fitted line is obtained as
\beqa
p_1&=& \frac{c_p}{\sqrt{|\Delta \mu/\epsilon|}} \exp \left[\frac{-g^*_p/2}{|\Delta \mu/\epsilon|-\Delta \mu_y(L)/\epsilon} \right],\nonumber \\
g^*_p &=& 0.423 \epsilon, \quad c_p=0.604. \label{eq_p1_black}
\eeqa
$\Delta \mu_y(L)$ for respective $L$ are shown in Table \ref{table_def}.
It should be noted that $g^*_p$ is very close to $g^*_{\rm Ising}$.

The lines for Eq.~(\ref{eq_p1_black}) with $L$ being $240\sqrt{2}a$ are shown by the dark solid lines in Fig.~\ref{p1T04}(b). 
The lines for Eq.~(\ref{eq_p1_black}) with $L$ being $160\sqrt{2}a$ and $80\sqrt{2}a$ are shown as pale solid lines in Fig.~\ref{p1T04}(b).
Although the only modification is the introduction of $\Delta \mu_y(L)$, the lines for Eq.~(\ref{eq_p1_black}) reproduce the $p_1$ based on the Monte Carlo results quite well for $\Delta \mu_{co}<|\Delta \mu|< \Delta \mu_R(L)$.

For large $|\Delta \mu|$, $p_1$ based on the Monte Carlo results depart from Eq.~(\ref{eq_p1_black}).
 The departing point $|\Delta \mu|/\epsilon=0.12$, agrees well with $\Delta \mu_R (240\sqrt{2}a)$ in Table \ref{table_def}.

For $\Delta \mu_y(L)< |\Delta \mu|$, the ``terrace'' has a slope $p_1$, so that the profile of the faceted macrostep becomes similar to that in the step droplet zone at equilibrium.
That is, the characteristic profile of the faceted macrostep in the step-faceting zone changes to that in the step droplet zone.
This $\Delta \mu_y(L)$ indicates a yielding point with respect to the self-detachment of steps from the macrostep.

It is interesting that $p_1$ is singular at the point $\Delta \mu_y(L)$.
$\Delta \mu_y(L)$ is a candidate for the non-equilibrium phase transition point.
 However, $\Delta \mu_y(L)< \Delta \mu_f(\tau,L)$ in the present study, so that the vicinal surface freezes around the yielding point. 
Therefore, phenomena regarding $\Delta \mu_y(L)$ were not realized.

\subsubsection{Step velocity \label{sec_stepV}}

%%%%% FIGURE  %%%%%

\begin{figure}%[h]%[htbp]
%\begin{center}
\centering
\includegraphics[width=7.0 cm,clip]{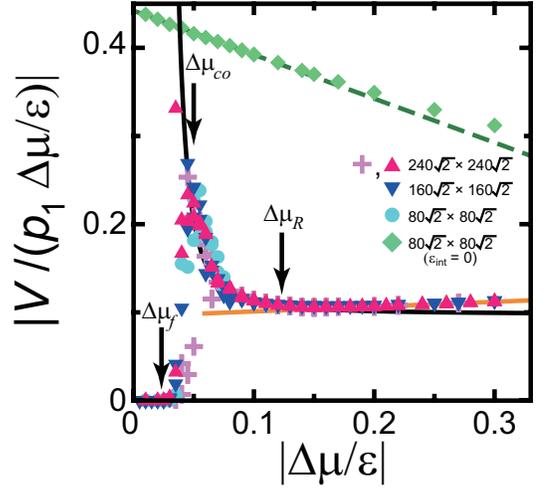}%
%\end{center}
\caption{
(Color online)
Absolute value of the step velocity $|v_{\rm step}| \equiv |(V/p_1)|$ divided by $|\Delta \mu /\epsilon|$.
Crosses indicate the Monte Carlo results obtained with the TS initial configuration.
Dark solid line: Eq.~(\ref{eq_vstep2}). 
Pale solid line: Eq.~(\ref{eq_vstep3}). 
Dashed line: Eq.~(\ref{eq_vstep4}).
The values were averaged over $2 \times 10^8$ MCS/site.
$\kBT/\epsilon=0.4$.
$\epsilon_{\rm int}/\epsilon=-0.9$.
}  
\label{VstepT04}
\end{figure}
%%%%% FIGURE  %%%%%

In the steady-state, $n_+=n_-=V/a$ and $n_+= p_1 v_{\rm step}/a$, where $a$ is the height of the elementary step.
Therefore, we obtain another key quantity $v_{\rm step}$ calculated with
\beq
v_{\rm step}=V/p_1   \label{eq_vstep}
\eeq
 using the Monte Carlo results for $V$ and $\langle n \rangle$. 
The step velocity $v_{\rm step}$ is approximately proportional to $\Delta \mu$ for $\Delta \mu_{co}(L)< |\Delta \mu|$.
Figure \ref{VstepT04} shows the $\Delta \mu$ dependence of $v_{\rm step}/|\Delta \mu/\epsilon|$.

$v_{\rm step}$ based on the Monte Carlo results can be fit by the following equations:
\beqa
v_{\rm step}/(\Delta \mu/\epsilon) &=& 0.094+3.2\times 10^{-3} \nonumber \\ 
&&\times \exp[0.18/|\Delta \mu/\epsilon|] \nonumber \\
&& \hspace{-1em} (\Delta \mu_{co}<|\Delta \mu/\epsilon|<\Delta \mu_R(L)),  \label{eq_vstep2}\\
v_{\rm step, RSOS}/(\Delta \mu/\epsilon) &=& 0.442-0.498|\Delta \mu|) \nonumber \\
&&\hspace{-1em} (\epsilon_{\rm int}=0, \  \text{RSOS model}).  \label{eq_vstep4}
\eeqa
These lines are shown in Fig.~\ref{VstepT04}.
The reason for the steep decrease in the step velocity as $|\Delta \mu|$ increases is the meeting of steps, which inhibits the growth/recession of steps substantially \cite{akutsu12}.
Figures \ref{surfdatT04}(d) and (e) show that the detached steps meet at several sites on the surface due to thermal fluctuations.
The sticky character of steps merges these steps locally.
The step velocity of the merged steps is substantially small \cite{akutsu12,akutsu14}, so that the merged steps pin the growth/recession of the steps.
The density of steps becomes larger as $|\Delta \mu|$ increases because $t_d$ becomes shorter.
The steps then meet and are pinned more frequently for large $|\Delta \mu|$. 
Therefore, the step velocity becomes smaller as $|\Delta \mu|$ increases.

It is interesting that the phases of the waves on the detached steps (meandering) are often coherent, although the surface does not contain dislocations or impurities.
Besides, the present model does not take the surface diffusion into account.
Nevertheless, the phases of the waves on the detached steps appear coherent.
This is because the advance/recession of the embryo formed at the edge of the macrostep is blocked more often by the preceding step when the location of the embryo is nearer to the unzipping point of the preceding step. 
The embryo formed at almost the center between the two unzipping points of the preceding step is more likely to survive.

For $|\Delta \mu| < \Delta \mu_{co}$, $l_d$ exceeds the linear size of the system, $L$ (Table {\ref{table_tn}}).
Therefore, the successive nucleation at the step edge breaks.
The surface moves intermittently through 2D heterogeneous nucleation at the step edge, which is consistent with the interpretation in the previous subsection (\S \ref{sec_2DV}).
In this region, $V/p_1$ does not indicate $v_{\rm step}$, because $n_-<v_{\rm step}p_1$.
$v_{\rm step}$ in this region should be the lesser of the two values obtained by Eqs.~(\ref{eq_vstep2}) and (\ref{eq_vstep4}).

\subsection{Consistency \label{sec_consistency}}

Using the equations for $p_1$ and $v_{\rm step}$ (Eqs.~(\ref{eq_p1_black})--(\ref{eq_vstep2})), we can reproduce $\langle n \rangle$ and $V$ for $\Delta \mu_{co}(L)<|\Delta \mu|<\Delta \mu_R(L)$.
$V$ is then expressed as follows:
\beqa
V&=& c_p \sqrt{|\Delta \mu/\epsilon|}\exp \left[\frac{-g^*_p/2}{|\Delta \mu/\epsilon|-\Delta \mu_y(L)/\epsilon} \right] \nonumber \\
&&\times \{0.094+3.2\times 10^{-3}  \exp[0.18/|\Delta \mu/\epsilon|]\}, \nonumber \\
&&g^*_p = 0.423 \epsilon, \quad c_p=0.604. \label{eq_Vsurf}
\eeqa
Equation (\ref{eq_Vsurf}) is shown in Figs.~\ref{GR} and \ref{lnyT04}(b), where 
the curves reproduce the Monte Carlo results well.

On the other hand, $\langle n \rangle$ is inversely expressed by $p_1$ from Eq.~(\ref{eq_p1}):
\beqa
\langle n \rangle &=&  \left (z + \frac{N_m}{N_{\rm step}}\right )^{-1}, \nonumber \\
z &=&  \frac{(\sqrt{2}-\bar{p})}{\bar{p}} \left ( \frac{\sqrt{2}}{p_1}-1 \right )^{-1}. %, 
 \label{eq_nav2}
\eeqa
In the limit of $p_1 \rightarrow 0$, $\langle n \rangle$ converges to $ N_{\rm step}/N_m$.
The $ N_{\rm step}/N_m$ values also reproduce the constant values of $\langle n \rangle$ for $\Delta \mu_f(\tau, L)< |\Delta \mu|< \Delta \mu_{co}(L)$.

In the case of $p_1 \neq 0$, $z$ is expressed by
\beqa
z^{-1} 
&=&\frac{\bar{p}}{(\sqrt{2}-\bar{p})} \left \{ \frac{\sqrt{2|\Delta \mu/\epsilon|}}{c_p}\exp \left[\frac{g^*_p/2}{|\Delta \mu/\epsilon|-\Delta \mu_y(L)/\epsilon} \right]\right.\nonumber \\
&&\left. -1 \right \} \label{eq_1/z} 
\eeqa
using Eqs.~(\ref{eq_p1_black}) and (\ref{eq_nav2}).
The lines of $\langle n \rangle$ are shown in Figs.~\ref{nav} and \ref{lnyT04}(a).
These lines also reproduce the Monte Carlo results well.

\subsection{Kinetic roughening \label{k-roughening}}

%%%%% FIGURE  %%%%%

\begin{figure}%[h]%[htbp]
%\begin{center}
\centering
\includegraphics[width=6.0 cm,clip]{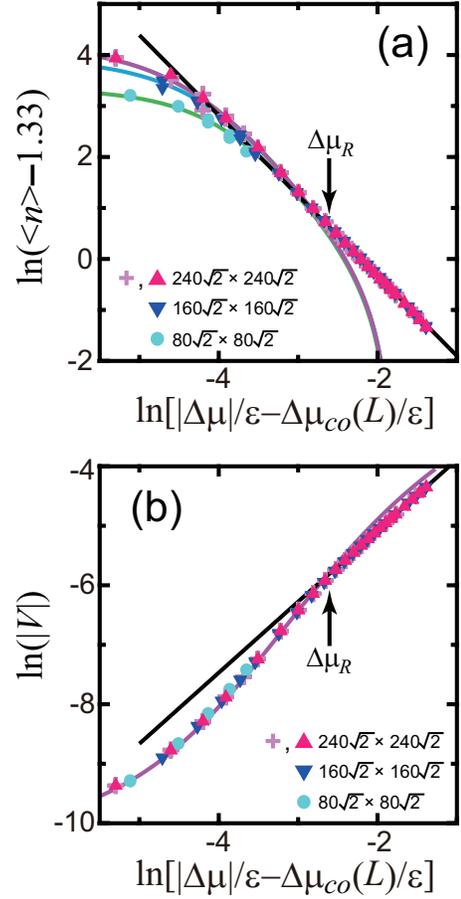}%
%\end{center}
\caption{
(Color online)
(a) Log-log plot of $\langle n \rangle$.
Dark solid line: Eq.~(\ref{eq_nav_power}).
Pale solid lines: Eqs.~(\ref{eq_nav2}) with (\ref{eq_1/z}) with $L=240\sqrt{2}$, $160\sqrt{2}$, and $80\sqrt{2}$ from the top to the bottom, respectively.
(b) Log-log plot of the absolute value of the surface growth rate.
Dark solid line: Eq.~(\ref{eq_V_power}).
Pale solid line: Eq.~(\ref{eq_Vsurf}).
Crosses indicate the Monte Carlo results obtained by the TS initial configuration.
$\kBT/\epsilon=0.4$.
$\epsilon_{\rm int}/\epsilon=-0.9$.
Averaged over $2 \times 10^8$ MCS/site.
}  
\label{lnxlnyT04}
\end{figure}
%%%%% FIGURE  %%%%%

For $\Delta \mu_R(L)<|\Delta \mu|$, the vicinal surface is kinetically roughened and the faceted macrostep disappears (Fig.~\ref{surfdatT04}(f)).
Although there is no large-scale macrostep, the inhomogeneous bumpy structure remains on the surface.
This bumpy structure is formed by thermal noise (Fig.~\ref{surfdatT04}(f)).

In this region, $\langle n \rangle$ and $V$ exhibit power law behavior (Fig.~\ref{lnxlnyT04}):
\beqa
\langle n \rangle &=& 0.0310(|\Delta \mu|/\epsilon -\Delta \mu_{co}(L)/\epsilon)^{-\zeta} +n_{\infty} \nonumber \\
\zeta &=&1.57\pm0.07, \quad n_{\infty}=1.33\pm0.08, 
\label{eq_nav_power} \\
|V|&=&0.0677(|\Delta \mu|/\epsilon -\Delta \mu_{co}(L)/\epsilon)^{\beta}, \nonumber \\ 
\beta &=& 1.19\pm0.05.\label{eq_V_power}
\eeqa
Here, the choice of $\beta$ as the symbol for the exponent is in accordance with Ref. \cite{depinning}.
The Monte Carlo results for all sizes agree with the two lines of Eq.~(\ref{eq_Vsurf}) or Eq.~(\ref{eq_V_power}), as shown in Fig.~\ref{lnxlnyT04}(b).
From the cross point of the two lines, $\Delta \mu_R(L)/\epsilon - \Delta \mu_{co} (L)$ is determined as $0.071\pm0.005$.
For $L=240\sqrt{2}a$, we have $\Delta \mu_R(240\sqrt{2}a)/\epsilon = 0.121\pm0.012$.

The step velocity is obtained from the Monte Carlo results in Fig.~\ref{VstepT04}.
The results are fitted to the following equation:
\beqa
v_{\rm step}/(\Delta \mu/\epsilon) &=& 0.096+ 0.055 |\Delta \mu/\epsilon| \nonumber \\
&&\hspace{-1em}  (\Delta \mu_R(L) < |\Delta \mu|),\label{eq_vstep3}
\eeqa
as represented by the pale solid line in Fig.~\ref{VstepT04}.
Using Eqs.~(\ref{eq_vstep3}) and (\ref{eq_V_power}), we obtain $p_1$ from $p_1=V/v_{\rm step}$: 
\beqa
p_1=\frac{0.0652(|\Delta \mu|/\epsilon -\Delta \mu_{co}/\epsilon)^{\beta}}{(\Delta \mu/\epsilon) (0.096+ 0.055 |\Delta \mu/\epsilon|),} \label{eq_p1power}
\eeqa
which is shown by pale solid lines in Fig.~\ref{p1T04}.
The line reproduces the Monte Carlo results for $p_1$ well.

The crossover from the vicinal surface with the faceted macrostep to the kinetically roughened surface is essentially caused by the change of the $|\Delta \mu|$ dependence of $p_1$ (Fig.~\ref{p1T04}(a)).
By definition (Appendix \ref{sec_p1-<n>}), $p_1$ indicates the density of the elementary steps.
The meeting of steps occurs more frequently when the step density is larger.
In Eq.~(\ref{eq_p1_black}), the merging of steps is not taken into consideration; 
Hence, the increase of the locally merged steps contributes to a decrease in $p_1$;  i.e., it changes the $|\Delta \mu|$ dependence of $p_1$.

Due to the locally merged steps, $|V|$ in the kinetically roughened region (Eq.~(\ref{eq_V_power})) becomes smaller than that expected from Eq.~(\ref{eq_Vsurf}) (Figs.~\ref{GR} and \ref{lnxlnyT04}(b)). 
In contrast, $\langle n \rangle$ in the kinetically roughened region (Eq.~(\ref{eq_nav_power})) becomes larger than that expected from Eq.~(\ref{eq_nav2}) (Figs.~\ref{lnyT04}(a) and \ref{lnxlnyT04}(a)).

\section{Discussion  \label{discussion}}

%tn <-> mu_f? 
%td <-> mu_y?
%ld <=> mu_co?

%L->infty: c-driving force?

Near equilibrium, the finite size effect is prominent.
The question then arises, what happens with the infinite system size?
The underline in the Table \ref{table_tn} shows the border value that exceeds the system size or the observation time in the present study.
From Eq.~(\ref{eq_tn}), the waiting time $t_n$ for 2D heterogeneous nucleation converges to zero as $L\rightarrow \infty$.
We then have $\lim_{L \rightarrow \infty} \Delta \mu_f (\tau,L)=0$.
However, $t_n$ increases so rapidly as $|\Delta \mu|$ decreases that the system size should be approximately $10^{14}$ or more for $t_n<10^8$ with $|\Delta \mu|/\epsilon=0.01$.
Therefore, in an actual system with a length of ca. 1 mm, the non-negligible frozen region with respect to the surface growth/recession remains near equilibrium.

For $l_d$ and $t_d$, there is no explicit size dependence.
The waiting time $t_d$, for the step detachment seems to be linked to $\Delta \mu_y(L)$.
The Monte Carlo results in this study show a slight $L$ dependence on $\Delta \mu_y(L)$.
If $\Delta \mu_y(L)$ converges to a non-zero value $\Delta \mu_y(\infty)$ in the limit of infinite system size, then the point may be a non-equilibrium phase transition point.
However, to clarify whether $\Delta \mu_y(\infty)$ is finite or zero is a future problem.

In our previous work \cite{akutsu17}, we studied the $\Delta \mu$ dependence of the size for faceted merged steps $\langle n \rangle$ and the growth rate of the vicinal surface $V$ in the step droplet zone I for the non-equilibrium steady-state.
Some results are similar to the results of this study.
$\langle n \rangle$ decreases as $|\Delta \mu|$ increases. 
In addition, for $|\Delta \mu|> \Delta \mu_R(L)$, the vicinal surface with faceted macrosteps crosses over to the kinetically roughened surface without a macrostep.
In the results of the previous study, the freezing region lacks a near equilibrium state, and the elementary step self-detaches by thermal noise without 2D nucleation processes.
The morphology of the kinetically roughened surface is somehow different from that of the present study.
The bumpy structures on the vicinal surface that were obtained in the previous study were so small that they cannot be discerned in the images of the simulated surface without magnification of the images \cite{akutsu17}. 
Therefore, the scaling behavior in the kinetically roughened surface is slightly different from that in the present study.

It is interesting that the figure for $V$ (Fig.~\ref{GR}) is analogous to the Fig.~4(b) in Ref. \cite{depinning}, 
which shows the velocity of the particles where plastic depinning occurs.
The system has a depinning threshold, $V \propto (F_D-F_c)^{\beta}$ with $\beta=1.5$, where $V$ is the average velocity of the particles, $F_D$ is the driving force, and $F_c$ is the depinning threshold.
The plasticity is said to be relevant to the charge-density wave systems \cite{coppersmith, depinning}. 
As noted in \S \ref{sec_stepV} and \S \ref{sec_consistency}, the elementary steps that detach from the faceted macrostep meet neighboring steps due to thermal noise.
The steps are sticky, so that they merge at the meeting point.
The locally merged steps with substantially low velocity then pin the motion of the elementary steps.
Therefore, the step attachment-detachment motion is analogous to the motion of particles with plastic depinning. 
This begs the question, is there is a common mathematical framework?
However, this question has yet to be answered.

To clearly elucidate the effect of the anomalous surface tension, we exclude surface diffusion \cite{uwaha92,sato2000}, volume diffusion \cite{nishinaga89, nishinaga93}, elastic interaction \cite{calogero, sutherland, alerhand, williams93, yamamoto94, song94, song95, dijken97, sudoh01, lossig96, bhattachariee,shenoy98, shenoy2000}, long range attractive interaction \cite{jayaprakash84-2, tonchev16}, and Ehrlich-Schwoebel effects \cite{ehrlich,schwoeble} from the model.
Nevertheless, the p-RSOS model in the step-faceting zone shows a wide variety of surface dynamics.
Combinations with the effect of the anomalous surface tension and other effects are future problems to be considered.

\section{Conclusions \label{conclusions}}
\begin{itemize}
\item Steps on the vicinal surface self-assemble to form faceted macrosteps in the steady-state for $|\Delta \mu|<\Delta \mu_R(L)$.
 
\item For $|\Delta \mu|< {\rm Max}[\Delta \mu_y(L), \Delta \mu_{f}(\tau, L)]$, the vicinal surface does not grow/recede.
 The ``terrace'' surface contacted with the (111) side surface of the faceted macrostep is the (001) surface (Fig.~\ref{surfdatT04}(a)).

\item For ${\rm Max}[\Delta \mu_y(L), \Delta \mu_{f}(\tau, L)]<|\Delta \mu|<\Delta \mu_{co}(L)$, the vicinal surface grows/recedes intermittently in the manner of classical 2D heterogeneous nucleation at the macrostep edge (Figs.~\ref{surfdatT04}(b) and (c)).  
 The critical size of the nucleus $l_c$ and the mean waiting time $t_n$ decrease as $|\Delta \mu|$ increases, according to Eqs.~(\ref{eq_size2D}) and (\ref{eq_tn}), respectively.
$V$ and $p_1$ are described by Eqs.~(\ref{eq_V_2D2}) and (\ref{eq_p1_black}), respectively. 
$\langle n \rangle \approx N_{\rm step}/N_m$ is constant, where $N_m=1.75$.
$v_{\rm step}$ is the smaller of Eqs.~(\ref{eq_vstep2}) and (\ref{eq_vstep4}).

\item For $\Delta \mu_{co}(L) < |\Delta \mu|< \Delta \mu_R(L)$, the vicinal surface grows/recedes in the manner of attachment-detachment of steps at the macrostep edge (Figs.~\ref{surfdatT04}(d) and (e), Fig.~\ref{multi_nuclei}).
The attachment-detachment of steps is understood based on successive classical 2D heterogeneous multi-nucleation at the edge of the macrostep.
The absolute value of the step velocity $|v_{\rm step}|$, the absolute value of the surface velocity $|V|$, and the slope of the ``terrace'' $p_1$ increase with $|\Delta \mu| $, according to Eqs.~(\ref{eq_vstep2}), (\ref{eq_Vsurf}), and (\ref{eq_p1_black}), respectively.
The characteristic length $l_d$, the step-detachment time $t_d$, and the average height of the merged step $\langle n \rangle$ decrease as $|\Delta \mu| $ increases, according to Eqs.~(\ref{eq_ld}), (\ref{eq_td}), and (\ref{eq_nav2}), respectively.

\item For $\Delta \mu_R(L) < |\Delta \mu|$, the vicinal surface roughens kinetically due to locally merged steps (Fig.~\ref{surfdatT04}(f)).
The surface quantities exhibit power law behavior; i.e., $|v_{\rm step}|$, $|V|$, and $p_1$ increase as $|\Delta \mu|$ increases according to Eqs.~(\ref{eq_vstep3}), (\ref{eq_V_power}), and (\ref{eq_p1power}), respectively.
In contrast, $\langle n \rangle$ decreases as $|\Delta \mu| $ increases, according to Eq.~(\ref{eq_nav_power}).

\item The finite size and the finite time effects are distinctive for $|\Delta \mu| \simless \Delta \mu_{co}(L)$.

\end{itemize}

\begin{acknowledgments}
The author acknowledges Prof. T. Sasada, Prof. T. Ohachi, Prof. T. Nishinaga, and Prof. V. Tonchev for continual encouragement.
This work is supported by Kakenhi Grants-in-Aid 
(Nos. JP25400413 and JP17K05503) from the Japan Society for
the Promotion of Science (JSPS).
\end{acknowledgments}

\appendix

\section{Calculation of the surface tension \label{surface-tension}}

To evaluate the surface free energy of the vicinal surface, the terms related to the Andreev field \cite{andreev} were added: $\vec{\eta}= (\eta_x, \eta_y)$. The Hamiltonian for the grand canonical ensemble with respect to the number of steps is \cite{akutsu98} 
\beqa
%&&
{\cal H}_{\rm vicinal}&=&{\cal H}_{\rm p-RSOS} %\nonumber \\
%   &&
-\eta_x \sum_{n,m}[h(n+1,m)-h(n,m)] \nonumber \\
   &&
-\eta_y \sum_{n,m}[h(n,m+1)-h(n,m)].
\label{hamil_vicinal}
\eeqa 
The Andreev field behaves similar to a chemical potential with respect to a single step.
The Legendre-transformed surface free energy introduced by Bhattacharjee \cite{bhattachariee} corresponds to the Andreev free energy \cite{andreev,akutsu15}.

%\subsection{Calculations based on statistical mechanics}

From a statistical mechanics perspective, the grand partition function ${\cal Z}$ is calculated as
%\beq
${\cal Z}= \sum_{\{h(m,n)\}} \exp[{-\beta {\cal H}_{\rm vicinal}}]$, %\label{partition}, 
%\eeq
 where $\beta= 1/\kBT$.
The summation with respect to $\{h(m,n)\}$ is taken over all possible values of $h(m,n)$.
The Andreev free energy $\tilde{f}(\vec{\eta})$ \cite{andreev} is the thermodynamic grand potential and is calculated from the grand partition function ${\cal Z}$ as \cite{akutsu98} 
\beq
\tilde{f}(\vec{\eta})= \tilde{f}(\eta_x, \eta_y)= - \lim_{{\cal N} \rightarrow \infty} \frac{1}{{\cal N}} \ \kBT \ln {\cal Z}, \label{tildef}
\eeq
where ${\cal N}$ is the number of lattice points on the square lattice.
$\vec{p}$ is also calculated using the PWFRG method from the equation $\vec{p} = (\langle h(m+1,n)-h(m,n)\rangle, \langle h(m,n+1)-h(m,n)\rangle)$. 

It should be noted that the profile of the Andreev free energy $\tilde{f}(\eta_x,\eta_y)$ is similar to the ECS $z=z(x,y)$, where 
$\tilde{f}(\eta_x,\eta_y) = \lambda z(x,y)$, $(\eta_x,\eta_y)=-\lambda( x, y)$, and $\lambda$ represents the Lagrange multiplier related to the crystal volume.

Using the inverse Legendre transform with respect to $\tilde{f}(\vec{\eta})$,  
\beq
f(\vec{p}) = \tilde{f}(\vec{\eta})+\vec{\eta}\cdot \vec{p}, \label{fpdef}
\eeq
we obtained the surface free energy $f(\vec{p})$ per unit $xy$ area.

\section{2D critical nucleus \label{2Dcritical}}

The excess free energy for an island that contacts the faceted macrostep $G(l)$ is expressed by Eq.~(\ref{eq_gibbs}).
For the critical nucleus, the Gibbs free energy is a minimum with respect to the shape, but is a maximum with respect to the size.
The shape of the critical nucleus is similar to the equilibrium island shape (Fig.~\ref{nuclei}(c)).
As for the size, we can derive Eq.~(\ref{eq_size2D}) in this Appendix.

If the length $l$ is replaced with $\lambda l$, where $\lambda$ is a scaling parameter, then we have $G(\lambda l)=-\lambda^2 \Delta \mu S(l) + \lambda \Gamma(l) - \lambda l \gamma_2^{(110)}$ for an island with a compact shape. 
Since $\rm{d}G(\lambda l)/\rm{d} \lambda =0$ at the critical nucleus, we have
\beqa
\left. \frac{dG(\lambda l_c)}{d \lambda}\right |_{\lambda \rightarrow 1} &=& [-2 \lambda \Delta \mu_1 S(l_c) +\Gamma(l_c) - l_c \gamma_2^{(110)}]|_{\lambda \rightarrow 1} \nonumber \\
&=& 0.
\eeqa
Then,
\beq
\frac{\Gamma(l_c) - l_c \gamma_2^{(110)}}{S(l_c)}=2 \Delta \mu_1.
\eeq
For different $\Delta \mu_2$, 
\beqa
\frac{(\Gamma(l_c) - l_c \gamma_2^{(110)})}{\lambda_2 S(l_c)}&=&2\Delta \mu_1/\lambda_2 \nonumber \\
&=&2 \Delta \mu_2.
\eeqa
Therefore, we have
$
\lambda_2=\Delta \mu_1/\Delta \mu_2.
$
$\Delta \mu$ can be selected arbitrarily, so that $\Delta \mu$ can be one, 
i.e., 
\beq
\lambda_2=\epsilon/\Delta \mu_2.
\eeq
If $l_{c,0}$ is calculated so that it satisfies
\beq
\Gamma(l_{c,0}) - l_{c,0} \gamma_2^{(110)}=2\epsilon S(l_{c,0}), \label{eq_Gamma0}
\eeq
then the Gibbs free energy is obtained by
\beq
G(l_{c,0})=(\Gamma(l_{c,0}) - l_{c,0} \gamma_2^{(110)})/2. \label{eq_G0}
\eeq
Therefore, for $\Delta \mu$, we have $l_c=l_{c,0}\epsilon/|\Delta \mu|$ and $G(l_c)=G(l_{c,0})\epsilon/\Delta \mu$.

\section{Relationship between $p_1$ and $\langle n \rangle$ \label{sec_p1-<n>}}

Let us consider a vicinal surface with the configuration of Figs.~\ref{surfdatT04}(b)--(d).
$\langle n \rangle$ is then approximated as
\beq
\langle n \rangle \approx N_{\rm step}/(N_1+ N_m), \label{eq_nav1}
\eeq
where $N_1$ is the number of single steps on the surface in contact with the (111) surface, and $N_m$ is the number of macrosteps. 
At the temperature $\kBT/\epsilon=0.4$, we assume $N_m \approx 1.75$.
Next, let us introduce $z$ and $x$ so that 
\beqa
N_1&=& z N_{\rm step}, \ N_{\rm macro}=(1-z)N_{\rm step} \nonumber \\
L_1&=&xL, \ L_{\rm macro}=(1-x)L, \label{def_z}
\eeqa
where $N_{\rm macro}$ is the number of elementary steps that compose macrosteps, $L_1$ is the linear length of the ``terrace'', and $L_{\rm macro}$ is the linear length of the macrosteps.
$N_1=z\bar{p}L/a=xp_1L/d$, so that
\beq
z\bar{p}=p_1 x, \label{z-x1}
\eeq 
and $N_{\rm macro}=(1-z)\bar{p}L/a=\sqrt{2}(1-x)L/a$, so that
\beq
1-x=(1-z)\bar{p}/\sqrt{2}. \label{z-x2}
\eeq
From Eqs.~(\ref{eq_nav1})--(\ref{z-x2}), we obtain Eq.~(\ref{eq_p1}).

%\nocite{*}

%\section{reference}

\begin{thebibliography}{999}
% Reference 1




\bibitem{mitani} Takeshi Mitani, Naoyoshi Komatsu, Tetsuo Takahashi, Tomohisa Kato, Shunta Harada, Toru Ujihara, Yuji Matsumoto, Kazuhisa Kurashige, and Hajime Okumura, ``Effect of aluminum addition on the surface step morphology of 4H--SiC grown from Si--Cr--C solution'', J. Cryst. Growth, {\bf 423}, 45 --49 (2015).


%TDGL
\bibitem{burkhardt} H. M\"{u}ller-Krumbhaar,T. W. Burkhardt, and D. M. Kroll, ``A generalized kinetic equation for crystal growth'', J. Cryst. Growth, {\bf 38}, 13--22 (1977). 
\bibitem{enomoto} Y. Enomoto, K. Kawasaki, T. Ohta, and S. Ohta, ``Interface dynamics under the anisotropic surface tension'',  Phys. Lett. {\bf 107}, 319--323 (1985).
\bibitem{barabasi} A. L. Barabasi and H. E. Stanley,  {\it Fractal Concepts in Surface Growth} (Cambridge University Press, Cambridge, UK, 1995). 
%

\bibitem{saito78} Y. Saito, ``Self-Consistent Calculation of Statics and Dynamics of the Roughening Transition'', Z. Phys. B {\bf 32}, 75--82 (1978).

\bibitem{pimpinelli} A. Pimpinelli, J. Villain, {\it Physics of Crystal Growth} (Cambridge University Press, Cambridge, UK, 1998).
%

\bibitem{phase_field} W. J. Boettinger, J. A. Warren, J. A., C.  Beckermann, C., A. Karma,  "Phase-Field Simulation of Solidification", Annual Review of Materials Research. {\bf 32} 163 (2002). doi:10.1146/annurev.matsci.32.101901.155803.



%2Dnucleation
\bibitem{bcf} W. K. Burton, N. Cabrera, F. C. Frank, `` The growth of crystals and the equilibrium structure of their surfaces'', Philos. Trans. Roy. Soc. Lond. A {\bf 243}, 299--358 (1951).


%nucleation 
\bibitem{kashchiev} D. Kashchiev, {\it Nucleatoin, Basic Theory with Applications}, (Butterworth/Heinemann, Oxford, UK, 2000).
\bibitem{dubrovskii}  V. G. Dubrovskii, {\it Nucleation Theory and Growth of Nanostructures}, (Springer-Verlag, Berlin, Heidelberg, 2014). 





\bibitem{becker} Von R. Becker and W. D\"{o}ring, ``Kintische Behandlung der Keimbildung in \"{u}bers\"{a}ttigten D\"{a}mpfen'', Ann. Physik, {\bf 24}, 719--752 (1935).
\bibitem{ookawa}  A. Ookawa, {\it Crystal Growth} (Sy\={o}kab\={o}, Tokyo,  1977) in Japanese.





\bibitem{tanaka14} K. K. Tanaka, J. Diemand, R. Ang\'{e}lil, and H. Tanaka, ``Free energy of cluster formation and a new scaling relation for the nucleation rate'', The Journal of Chemical Physics {\bf 140}, 194310 (2014); doi: 10.1063/1.4875803.


\bibitem{kimura12} Y. Kimura, K. K. Tanaka, H. Miura, K. Tsukamoto, ``Direct observation of the homogeneous nucleation of manganese in the vapor phase and determination of surface free energy and sticking coefficient'', Crystal Growth \& Design, {\bf 12}  3278-3284 (2012).

\bibitem{kimura16} S. Ishizuka, Y. Kimura, T. Yamazaki, T. Hama, N. Watanabe, and A. Kouchi, ``Two-step process in Homogeneous Nucleation of Alumina in Supersaturated Vapor'', Chemistry of Materials, {\bf 28} 8732-8741 (2016).


\bibitem{nanev15} C. N. Nanev and V. D. Tonchev, ``Sigmoid kinetics of protein crystal nucleation'', J. Cryst. Growth {\bf 427} 48-53 (2015).



\bibitem{k_roughening} J. Krug, P. Meakin, ``Kinetic roughening of Laplacian fronts'', Phys. Rev. Lett. {\bf  66}, 703--706 (1991).

\bibitem{uwaha92} M. Uwaha, Y. Saito, ``Kinetic smoothing and roughening of a step with surface diffusion'', Phys. Rev. Lett. {\bf 68}, 224--227 (1992).



\bibitem{akutsu09} N. Akutsu, ``Thermal step bunching on the restricted solid-on-solid model with point contact inter-step attractions'', Applied Surface Science, {\bf 256}, 1205--1209 (2009). 
doi:10.1016/j.apsusc.2009.05.080.





\bibitem{akutsu11}
N. Akutsu, ``Zipping process on the step bunching in the vicinal
surface of the restricted solid-on-solid model with the step
attraction of the point contact type'', J. Cryst. Growth, {\bf 318}, 10--13 (2011). 
doi: 10.1016/j.jcrysgro.2010.10.088. 


\bibitem{akutsu11JPCM} N. Akutsu, `` Non-universal equilibrium crystal shape results from sticky steps'',  J. Phys. Condens. Matter, {\bf 23}, 485004 (2011). 
doi:10.1088/0953-8984/23/48/485004.

\bibitem{akutsu12} N. Akutsu, ``Sticky steps inhibit step motions near equilibrium'',  Phys. Rev. E {\bf 86}, 061604 (2012). 
DOI: 10.1103/PhysRevE.86.061604.

\bibitem{akutsu14} N. Akutsu, ``Pinning of steps near equilibrium without impurities, adsorbates, or dislocations'', J. Cryst. Growth, {\bf 401}, 72--77 (2014).
http://dx.doi.org/10.1016/j.jcrysgro.2014.01.068.

\bibitem{akutsu16} N. Akutsu, ``Faceting diagram for sticky steps'',  AIP Adv., {\bf 6}, 035301 (2016).   dx.doi.org/10.1063/1.4943400.

\bibitem{akutsu16-2} N. Akutsu, ``Effect of the roughening transition on the vicinal surface in the step droplet zone'',  J. Cryst. Growth {\bf 468}, 57--62 (2017). doi:10.1016/j.jcrysgro.2016.10.014.


\bibitem{akutsu16-3} N. Akutsu, ``Profile of a Faceted Macrostep Caused by Anomalous Surface Tension'',  Adv. Condens. Matter Phys. {\bf 2017}, Article ID 2021510 (2017). doi:10.1155/2017/2021510.



\bibitem{cabrera} 
N. Cabrera, R. V. Coleman, {\it The Art and Science of Growing Crystals},  Ed. J. J. Gilman, (John Wiley \& Sons: New~York,  London, 1963). 
\bibitem{cabrera64}
N. Cabrera,  ``The equilibrium of crystal surfaces'', Surf. Sci.  {\bf  2}, 320--345 (1964). 





%Mermin-Wagner theorem
\bibitem{mermin}  N. D. Mermin, H. Wagner, `` Absence of ferromagnetism or antiferromagnetism in one-or two-dimensional isotropic Heisenberg models'',  Phys. Rev. Lett.  {\bf 1966}, {\bf 17}, 1133--1136 (1966). 
 ``Erratum: Absence of ferromagnetism or antiferromagnetism in one-or two-dimensional isotropic Heisenberg models. N. D. Mermin and H. Wagner [Phys. Rev. Letters 17, 1133 (1966)].'' 
 Phys. Rev. Lett., {\bf 17}, 1307 (1966). 


\bibitem{dmrg} S. R. White, ``Density matrix formulation for quantum renormalization groups'', Phys. Rev. Lett., {\bf 69}, 2863--2866 (1992). 
\bibitem{dmrg2} T. Nishino, ``Density matrix renormalization group method for 2D classical models'',  J. Phys. Soc. Jpn., {\bf 64}, 3598--3601 (1995).
\bibitem{dmrg3} U. Schollw\"{o}ck, ``The density-matrix renormalization group, Rev.~Mod. Phys., {\bf 77}, 259--315 (2005). %


%
\bibitem{pwfrg} T. Nishino, K. Okunishi, ``Product wave function renormalization group'', J. Phys. Soc. Jpn., {\bf 64}, 4084--4087 (1995). 
\bibitem{pwfrg2} Y. Hieida, K. Okunishi, and Y. Akutsu, ``Magnetization process of a one-dimensional quantum antiferromagnet: The product-wave-function renormalization group approach'', Phys. Lett. A, {\bf 233}, 464--470 (1997). 
\bibitem{pwfrg3} Y. Hieida, K. Okunishi, and Y. Akutsu, ``Numerical renormalization approach to two-dimensional quantum antiferromagnets with valence-bond-solid type ground state'', New J. Phys., {\bf 1}, 7 (1999).
%


\bibitem{gmpt} E. E. Gruber, W. W. Mullins, ``On the theory of anisotropy of crystalline surface tension'', J. Phys. Chem. Solids {\bf 28}, 875-887 (1967).
\bibitem{gmpt2}
V. L. Pokrovsky, A. L. Talapov, ``Ground state, spectrum, and phase diagram of two-dimensional incommensurate crystals'', Phys. Rev. Lett. {\bf 42}, 65--67 (1979).
%
\bibitem{beijeren87} Van H. Beijeren, I. Nolden, `` Structure and Dynamics of Surfaces'', Eds. W. Schommers,  von P. Blancken-Hagen, (Springer,  Berlin, Heidelberg, Germany, 1987) Volume 2, p. 259.



\bibitem{sato2000} M. Sato, M. Uwaha and Y. Saito, ``Instabilities of steps induced by the drift of adatoms and effect of the step permeability'', Phys. Rev. B {\bf 62}, 8452--8472 (2000).


\bibitem{nishinaga89} T. Nishinaga, C. Sasaoka, and A. A. Chernov, `` A numerical analysis for the supersaturation distribution around LPE macrostep'', Ed. I. Sunagawa, {\it Morphology and Growth Unit of Crystals} (Terra Scientific Publishing Company, Tokyo, 1989).
\bibitem{nishinaga93} T. Nishinaga and T. Suzuki, ``Towards understanding the growth mechanism of III-V semiconductors on an atomic scale'', J. Cryst. Growth, {\bf 128} 37-43 (1993).



%g/l^2
\bibitem {calogero} F. Calogero, ``Solution of a threebody problem in one dimension'', J. Math. Phys., {\bf 10}, 2191--2196 (1969).
\bibitem{sutherland}
B. Sutherland, ``Quantum Many Body Problem in One Dimension: Ground State'', J. Math. Phys., {\bf 12}, 246--250 (1971).

\bibitem{alerhand} O. L. Alerhand, D. Vanderbilt, R. D.  Meade, and J. D.  Joannopoulos, ``Spontaneous formation of stress domains on crystal surfaces'', Phys. Rev. Lett., {\bf 61}, 1973--1976 (1988).
%
\bibitem{williams93} E. D. Williams, R. J. Phaneuf, J. Wei, N. C. Bartelt, and T. L. Einstein, ``Thermodynamics and statistical mechanics of the faceting of stepped Si (111)'', Surf. Sci. {\bf 294}, 219--242 (1993).
 Erratum to ``Thermodynamics and statistical mechanics of the faceting of stepped Si (111)'' [ Surf. Sci. 294 (1993) 219], Surf. Sci. {\bf 310},~451 (1994).
%

\bibitem{yamamoto94} T. Yamamoto, Y. Akutsu, and N. Akutsu, `` N. Fluctuation of a single step on the vicinal surface-universal and non-universal behaviors'', J. Phys. Soc. Jpn. {\bf 63}, 915--925 (1994).



%Si(113)
\bibitem{song94} S. Song and S. G. J. Mochrie, `` Tricriticality in the orientational phase diagram of stepped Si (113) surfaces'', Phys. Rev. Lett., {\bf 73}, 995--998 (1994). 
\bibitem{song95} S. Song and S. G. J. Mochrie, `` Attractive step-step interactions, tricriticality, and faceting in the orientational phase diagram of silicon surfaces between [113] and [114]'', Phys. Rev. B, {\bf 51}, 10068--10084 (1995).
%
\bibitem{dijken97} S. van Dijken, H. J. W. Zandvliet, and B. Poelsema, ``Anomalous strong repulsive step-step interaction on slightly misoriented Si (113)'', Phys. Rev. B {\bf \em55}, 7864--7867 (1997).
%%
\bibitem{sudoh01} K. Sudoh and H. Iwasaki, ``Kinetics of Faceting Driven by Attractive Step-Step Interactions on Vicinal Si(113)'', Phys. Rev. Let. {\bf 87}, 216103 (2001). 
%
%
\bibitem{lossig96} M. Lassig, ``Vicinal surfaces and the Calogero-Sutherland model'', Phys. Rev. Lett. {\bf  77}, 526--529 (1996).
%
\bibitem{bhattachariee} S. M. Bhattacharjee, ``Theory of tricriticality for miscut surfaces'',  Phys. Rev. Lett. {\bf 76}, 4568--4571 (1996).


\bibitem{shenoy98} V. B. Shenoy, S. Zhang, W. F. Saam, ``Bunching transitions on vicinal surfaces and quantum n-mers'', Phys. Rev. Lett., {\bf 81}, 3475--3478 (1998).
\bibitem{shenoy2000}  V. B. Shenoy, S. Zhang, W. F. Saam, ``Step-bunching transitions on vicinal surfaces with attractive step interactions'', Surf. Sci., {\bf 467}, 58--84 (2000).



\bibitem{jayaprakash84-2} 
C. Jayaprakash, C. Rottman, and W. F. Saam, ``Simple model for crystal shapes: Step-step interactions and facet edges'', Phys. Rev. B {\bf 30}, 6549--6554 (1984). 
%
\bibitem{tonchev16} A. Krasteva, H. Popova,N. Akutsu, and V. Tonchev, ``Time scaling relations for step bunches from models with step-step attractions (B1-type models)'', AIP Conf. Proc., {\bf 1722}, 220015 (2016). doi:10.1063/1.4944247.



\bibitem{ehrlich}  G. Ehrlich and F. G. Hudda, ``Atomic View of Surface Self?]Diffusion: Tungsten on Tungsten'', J. Chem. Phys. {\bf 44} 1039 (1966). 
\bibitem{schwoeble} R. L. Schwoeble and E. J. Shipsey, ``Step Motion on Crystal Surfaces '', J. Appl. Phys. {\bf 37} 3682 (1966).

\bibitem{osman} S. M. Osman, B. Grosdidier, I. Ali, and A. B. Abdellah, ``Liquid gallium-lead mixture phase diagram, surface tension near the critical mixing point, and prewetteing transition'', Phys. Rev. E {\bf 87}, 062103 (2013).




\bibitem{akutsu87} N. Akutsu, Y. Akutsu, ``Roughening, faceting and equilibrium shape of two-dimensional anisotropic interface. I. Thermodynamics of interface fluctuations and geometry of equilibrium crystal shape'',  J. Phys. Soc. Jpn.,{\bf 56}, 1443--1453 (1987).


%ecs
\bibitem{andreev} A. F. Andreev, ``Faceting phase transitions of crystals'', Sov. Phys. JETP, {\bf 53}, 1063--1069 (1981).

\bibitem{landau} L. D. Landau, E. M. Lifshitz, {\it Statistical Physics}, 2nd ed. (Pergamon,  Oxford, UK, 1968). 




%wulff theorem
\bibitem{laue} Von M. Laue, ``Der Wulffsche Satz f\"{u}r die Gleichgewichtsform von Kristallen'', Z. Kristallogr. {\bf 105}, 124--133 (1944).
\bibitem{herring} C. Herring, ``Some theorems on the free energies of crystal surfaces'', Phys. Rev. {\bf 82}, 87--93 (1951).
\bibitem{mackenzie} J. K. MacKenzie, A. J. W. Moore, and J. F. Nicholas, ``Bonds broken at atomically flat crystal surfaces--I: face-centred and body-centred cubic crystals'', J. Chem. Phys. Solids {\bf 23}, 185--196 (1962).










\bibitem{akutsu86} Y. Akutsu and N. Akutsu, ``Relationship between the anisotropic interface tension, the scaled interface width and the equilibrium shape in two dimensions", J. Phys. {\bf A19} (1986) 2813--2820.
\bibitem{akutsu90} Y. Akutsu and N. Akutsu, "Interface tension, equilibrium crystal shape, and imaginary zeros of partition function: Planar Ising systems", Phys. Rev. Lett. {\bf 64}, 1189 (1990).
\bibitem{akutsu92} N. Akutsu, ``Equilibrium Crystal Shape of Planar Ising  Antiferromagnets in External Fields'', J. Phys. Soc. Jpn. {\bf 61} (1992) 477.



\bibitem{akutsu17} N. Akutsu, "Disassembly of Faceted Macrosteps in the Step Droplet Zone in Non-Equilibrium Steady State", Crystals, {\bf 7}, Article ID cryst7020042, (2017).  doi:10.3390/cryst7020042.


\bibitem{depinning} C. Reichhardt, C. J. Olson Reichhardt, ``Depinning and nonequilibrium dynamic phases of particle assemblies driven over random and ordered substrates: A review'', Reports on Progress in Physics {\bf  80}, 026501 (2017).


\bibitem{coppersmith} S. N. Coppersmith, ``Phase slips and the instability of the Fukuyama-Lee-Rice model of charge-density waves'', Phys. Rev. Lett. {\bf 65} 1044 (1990). 


\bibitem{akutsu98} N. Akutsu,Y. Akutsu, ``Thermal evolution of step stiffness on the Si (001) surface: Temperature-rescaled Ising-model approach'', Phys. Rev. B, {\bf 57}, R4233--R4236 (1998).


\bibitem{akutsu15} N. Akutsu and T. Yamamoto, ``Rough-Smooth Transition of Step and Surface'', {\it Handbook of Crystal Growth}, Volume I, p. 265.  Ed. T. Nishinaga, (Elsevier, Amsterdam, Boston, Heidelberg, London, Paris, Singapopore, Tokyo,  2015) .









\end{thebibliography}
%\bibliographystyle{mdpi}

%=====================================
% References, variant A: internal bibliography
%=====================================
%\renewcommand\bibname{References}

%=====================================

%\section{end}

\end{document}